\documentclass[epj]{svjour}
\usepackage{graphics}
\begin{document}
\title{Entropy Production and Particle Yields in Heavy Ion Collisions at LHC}
\author{B\'ela Luk\'acs \and Andr\'as Ster
}                     
\offprints{ster.andras@wigner.mta.hu}          
\institute{MTA KFKI WFI, H-1525 Budapest 114, PO Box 49, Hungary}
\date{Received: date / Revised version: date}
%
\abstract{
A scheme is proposed for calculating the entropy production
in highly transparent colliding systems. The formalism is continuum
physical and fully conform with thermodynamics; and the calculation
is explicit in entropy production.
We have analyzed high energy heavy ion collisions from viewpoint
of thermodynamics. Entropy densities are calculated for reactions up
5.52 TeV/nucleon-pair energies. Final state particle productions are predicted using
particle generating quark and hadron models. We find that at LHC
energy the entropy density is 65.7. From this value we could
predict particle yields. The results show that there is not yet asymptotic freedom.
} 
\maketitle
\section{Introduction}
\label{intro}

A group of physicists in Budapest recognised in 1987 that calculations of heavy ion 
collision events need i) thermodynamic approach because of the lack of any a priori 
knowledge about rehadronisation processes; and ii) the development of a self-consistent 
anisotropic continuum dynamics proper for numeric
calculations, because of transparency in ultrarelativistic collisions 
\cite{luk84,luk87,plos,apsl,luk900,luk901,luk902,luk903,luk91,zim93,arv91,kfki93,luk96,zim00,holba,luk86,luk89}.

Such an approach can be used only once before the start of a new accelerator, 
of course, and then maybe once more when the first well-established yield ratios 
have been known, to nail down the actual rehadronisation model at that energy.

In this paper we want to give a prediction for the hadronic yields at LHC in Pb+Pb collisions
with maximal center of mass energy of 5.52 TeV/nucleon-pair. The apparatus will
work at full energy and intensity in 2014; and chances are substantial that even our
predictions may turn out to be wrong. However, in this case that would help, as well.

The hadronic yields depend on many factors. We know some of them fairly well. Some we
believe to know fairly well. And we are sure that we do not know some well. For example,
we more or less think that we can predict the specific entropy (S/N) at maximal temperature well.
We believe that the state of Quark-Gluon Plasma (QGP) is described correctly.
However, we know that we cannot
select the correct rehadronisation process without empirical experience.

In any case, we do not yet have this experience at 5.52 TeV/nucleon-pair. So the strategy will be as follows.
We use:

1) A dynamic calculation up to maximal temperature, so obtaining the specific entropy.

2) Then we switch to QGP, exploiting thermodynamics.

3) The blob starts to expand and cool, so rehadronisation comes,
for which we use 10 different models, + one calculation without QGP phase.

4) The rehadronisation models give various yields. For SPS and RHIC energies
(158 \& 200 AGeV fixed target, 130 \& 200 AGeV storage ring) they can be compared to measurements,
whence we extrapolate for the best rehadronisation model
at LHC energy (at least within the space spanned by the present models).

The rehadronisation models used in this paper are listed and briefly explained 
in Appendix~\ref{app:mod}.

Thus we will give our most preferred prediction. Also we will give a few other ones for cases
when some factor went against our best efforts. E.g. the dynamic calculation may be proven
invalid (e.g. because of deviation from the extrapolation of the differential cross sections).
If so the prediction is given for another alternative specific entropy, as well; and so on.

This way, if our present most preferred prediction proves approximately correct in some months,
then we can conclude that nothing unexpected happened. If, in contrast, one of the alternative
predictions is better, then we at least can guess what unexpected happened. Of course, it is possible
that neither our preferred predictions nor any of the alternative ones will be good; then we will have
been proven brave but unsuccessful.

In Section~\ref{chap:his} we give a bird's eye view of thermodynamic rehadronisation models,
while Section  \ref{chap:form} calculates the specific entropy S/N produced in the collision at
SPS, RHIC and LHC energies, with flows highly anisotropic because of the
substantial transparency. Section~\ref{chap:numr} gives the respective numeric values for S/N.

In Section~\ref{chap:app} we select a finite set of rehadronisation models from among the infinite
possibilities. Which of them 8 will be based on the probabilities of quark encounters, with
or without, however, the inclusions of compressibility, influence of masses of resulting
hadrons and/or gluon fragmentation. One more employs sequential fission as rehadronisation
process, and the last one, similar to Model 7, will be defined in due course.
In addition, for comparison, we will use a model without QGP, as well.

Of course, we cannot know a priori, which model is the best approximation, moreover,
this may depend on the accelerator energy, as well. So we calculate the hadronic yields
for the previous SPS \& RHIC energies, and then we can get an educated guess for LHC.

This way we select our best candidate, together with 2 less probable ones (and we
repeat the calculations for the absence of QGP, for comparison). We definitely do have a
unique "best candidate"; the success of one of the alternates would mean that some unexpected
happened between 1 and 5.52 TeV.

Section~\ref{chap:dis} discusses some simplifications we deliberately have done, and tries to give
corrections. Section~\ref{chap:con} gives the conclusions.

Some of the heavy ion physicists in Budapst from time to time give predictions for future 
accelerators or thermodynamic comments to preliminary results back to 1987; 
references will come in due course.
The tasks are taken with gaps of many years; this is the first prediction from us
for LHC, assuming that the thermodynamic approach works at least moderately well
in this regime, too.

\section{On the history of thermodynamic predictions of rehadronisation}
\label{chap:his}

Since free quarks cannot be produced, heavy ion physics always was regarded as the
only experimental way to produce Quark-Gluon Plasma. Back to 1979 the predictions
were that for this either several normal nuclear densities or high temperatures (cca. $>$ 160 MeV)
are needed. Because heavy ion collisions produce high density via collisions, heating up
is automatic. Therefore, generally predictions tended to be optimistic: roughly always:
"just the next accelerator will produce QGP".

Of course, the Budapest group was not immune from this optimism. In the middle of the '80's,
AGS being constructed at BNL we guessed that 9 GeV/A beam energy (on fixed target) would be
enough to get pure QGP~\cite{luk84}. We were not alone with this expectation. True, in 1987
the thermodynamic investigations drew attention to the fact that the slow decay of
longitudinal momentum may cause a low-density start of the phase transition~\cite{luk87} which, however,
cannot end for a while. AGS started in 1986, and after preliminary results we believed in QGP
at 14.5 GeV/A Si+Pb~\cite{plos}; we believed the high $K^+$/$\pi^+$ ratio a signal that the
$K^+$ came from QGP.
The subsequent data confirmed the high ratio~\cite{tann}, but a methodical comparative analysis
confronting QGP \& Hadronic Matter (HM) scenarios, including dynamic calculations of both collision and new
phase formation would have been necessary but then partially inavailable.
For example, we performed a calculation taking finite nucleation time in the phase transition
into account~\cite{apsl}, but we did not know, of course, the nucleation time; and the highly
anisotropic local state made hydrodynamics unfounded.
So our next analysis could have gone only to the breakup~\cite{luk900}, we only concluded about
a quite normal hadronic breakup,
with unknown prehistory. It became then obvious that dynamic calculations of the collisions are
inevitable: for this we elaborated the equations~\cite{luk901,luk902}, and performed some 
hydrodynamic calculations for the anisotropic matter formed in the collision~\cite{luk903}.
Since the result was a
very extensive transparency preventing too much density increase, the QGP formation seemed more
and more improbable at 9 GeV/A (fixed target).

Here it is necessary to stop for a moment. Hadrons cannot carry information about QGP; leptons in
principle can, but for them the conclusions are not easy. We detect the hadrons from the
Hadronic Matter,
after rehadronisation. Now, the turbulent and unfamiliar state of matter at maximal
compression may
behave as Quark Phase from one aspect, and as Hadronic Phase from another. Also, very probably,
the phase transition is of first order, and then, even in equilibrium, mixed phase is possible,
in which case we may not recognise what happened really. It is no surprise that contradictory
conclusions were drawn from AGS experiments.

From 1991 attention turned to the SPS under construction, at energies more than 1 order of
magnitude higher then AGS (and to the future RHIC). It was not clear if the higher energy
would
be enough to get QGP; therefore, we performed a multimodel prediction, with hadronic matter phase
as well as QGP with sequential fission hadronisation and with various nonequilibrium phase
transitions, as well. (This model family will be used in the present paper too.) For the scheme
see \cite{luk91,zim93,arv91,kfki93,luk96}.

When the SPS started to give data, worldwide expectations were equivocal, but somewhat in favour
of QGP. However, after 1997 it was generally believed that no well developed QGP was present
(see e.g.~\cite{luk96,let96,let93}), even if some doubts lingered. In 1999 the model ALCOR
(Algebraic Coalescence Rehadronisation)
was elaborated \cite{zim00}, where quarks are deliberated but
do not form a genuine phase. While ALCOR itself is not explicitly part of our multimodel scenario
(being the scenario formed back in 1991), its philosophy is similar to Sequential Fission, which is our
Model 10. ALCOR was able to reproduce some SPS yield ratios, and we think now that it would
have been
successful even earlier to explain the high $K^+$ yields \&c., unexplained in pure Hadronic Matter
scenarios.

Then came the RHIC in 2000-1, and practically everybody expected QGP. However, \cite{zim00} was a
warning that from yield ratios the signature may be unclear. And here observe that we simply cannot
make any thermodynamically well-founded statement about the transient presence of a QGP phase at
maximal temperature. For that we should prepare and observe systems with different choices of
independent extensives~\cite{lan61}, now particle components, so deciding if the prepared systems
are similar to each other, but if one cannot do this, then  the programme cannot be realised.
The liberated quarks guessed in 1999
in~\cite{zim00} may quite have been present, but we cannot observe it. However, we can observe
the hadronic yields.

From that time (2003) there is a study comparing various scenarios when combination/recombination
happens before detection, so we cannot observe the combination processes in themselves. Two of the
cases were primordial hadronisation and heavy ion rehadronisation, for the second Models 1-10 were
discussed upto S/N=50~\cite{holba}.

\hyphenation{aniso-tropic}

We are before the start of data about the yields at 7000+7000 AGeV storage ring energies
for p+p and at the corresponding 2760+2760 AGeV for Pb+Pb collisions.
No doubt, the collision energies will be high enough for QGP. However, extreme transparency means
that high densities are not expected; high temperatures may be present, but to decide this the collision
must be followed up in some anisotropic continuum scenario. As for signatures, no doubt,
at the same temperature
QGP may and would result in yields very different from pure Hadronic Matter;
but the yields are expected to depend strongly not only on the  transient presence of QGP but
also on the actually working rehadronisation process; we do not have a priori information about
the process, and practically the only way to decide the dominant way of the rehadronisation is to
deduce it from the yields. However, we think that now we are in a position to make definite
predictions.

Namely, it is almost consensus that QGP was present already in the RHIC experiments. So some of
our rehadronisation models with QGP, or at least some combinations of them, must give tolerable
yields at 65+65 \& 100+100 AGeV, and then we can guess the best model for 2760+2760 AGeV.
This will be our method in this paper.

\section{Formulae of entropy production}
\label{chap:form}

\subsection{On the anisotropic state}

For simplicity throughout the whole paper we consider a matter composed of a
single, conserved particle component. This restriction is not necessary at all,
could be lifted, but now strongly simplifies the equations. This conserved
component is, of course, the baryons.

This only component has a flux vector, $n^i$. Except for very exotic, probably
unphysical situations this flux vector is timelike (a spacelike flux would
mean acausal motion for the particles, and a lightlike one would be a limiting
situation in the best case). It is algebraically clear that a timelike flux
vector field always can be decomposed as
\begin{equation}
        n^i  \equiv nu^i; \,\,\,\,\,  u_ru^r = 1
\label{eq:flux}
\end{equation}
where $u^i$ is the velocity field of the continuum and $n$ is the comoving
particle number density. (According to the Einstein convention of General Relativity,
if an index appears pairwise, above and below, then summation is automatically meant.)
In local equilibrium the velocity field $u^i$ is not only the average of the individual
particle velocities, but also a significant part of the particles have velocities
near to $u^i$. However, in the present case of very energetic collisions the two original
nuclei do not stop each other but they interpenetrate. Then the resulted momentum
distribution is not even similar to a "thermal" one, but the momenta bins about  $m*u^i$ are almost
depopulated. This is a nontrivial, challenging situation for Thermodynamics.

Almost any thermodynamic description is possible in 2 conventions. Either energy E
is the thermodynamic potential, the function of the (independent) extensives, of
which entropy S is one, or S is the potential, and E is one of the extensives.
The two conventions are almost equivalent, and again the exotic details have
nothing to do with anisotropy. So just for definiteness' sake, let the potential be S.

Then S must be a unique function of the set of independent extensives, moreover a
homogeneous linear function,
\begin{equation}
        S = Y_RX^R = S,_RX^R
\label{eq:Sr}
\end{equation}
(beware the Einstein convention). Without (2) there would be no Thermodynamics at all.

Now, which are the independent extensives? We cannot know the answer a priori~\cite{lan61}.
Indeed, there are cases when even the number of independent particle components is
not obvious~\cite{luk86}. While that is 1 in this Section, the correct way is always
first to take a set of extensives which seems to be minimal; and if that is not
sufficient, we introduce one more and so on.

In Ref.~\cite{luk87} we manufactured the simplest extension for anisotropic local states,
with an extra extensive Q. In the simplest case Q has a dimension momentum*volume,
so Q/V is momentum (or, alternately, Q is the momentum*particle number), and if Q=0 then
the local state is isotropic. So for order of magnitude we expect Q/N to be similar
to the measure of momentum anisotropy.

At this point we chose simple nuclear and quark equations of state, and we used some
Fundamental Laws of Thermodynamics and so. The result was: at 0 anisotropy the
deconfinement phase transition starts at 5 times the normal nuclear densities and ends at
11 times the normal nuclear densities. In between there is an n-q mixture. With increasing
anisotropy the phase transition starts at lower and lower densities, but ends
at higher and higher ones.

Since we do not know how big the anisotropy will be at full overlap, generally
we cannot avoid detailed dynamic calculations, but the approach does not need
the calculation of distributions. Anyway, as transparency increases, anisotropy increases
as well, so nobody should be much surprised if any possible experiment would end
with n-q mixtures.

Ref.~\cite{luk87} used the simplest nuclear equations of state, and a perturbative QCD one for
the quark plasma. In~\cite{luk89} we made the anisotropic equations of state for Skyrme potential;
not as if we believed it but as an explicitly momentum-dependent one. If there is a
favourite potential + nice arguments, we can repeat the construction in~\cite{luk89}
for that one. Now let us focus on entropy production.

\subsection{Dynamics}

We have the more complicated task: an anisotropic, strongly interacting,
transparent system, starting from maximally anisotropic but cold state,
and the question is (more or less): what will be/was present at maximal density?
(The observation is at asymptotic outgoing state, so well after maximal
density).

But we have 2 undisputable equations and 1 undisputable unequality:
\begin{equation}
  T^{ir}_{;r} = 0
\label{eq:T0}
\end{equation}
\begin{equation}
  n^r_{;r} = 0
\label{eq:n0}
\end{equation}
\begin{equation}
  s^r_{;r} \geq 0
\label{eq:s0}
\end{equation}
The derivations of the dynamic+thermodynamic equations are given in~\cite{luk901};
here we give only Schlagw\"orter. Everything has an energy-momentum tensor T
(see GR); the system is compact and separated, so Eq.~(\ref{eq:T0}) is true.
Eq.~(\ref{eq:n0}) is true for conserved particles; if they are not conserved, write a rhs.
And Eq.~(\ref{eq:s0}) is the Second Law of Thermodynamics. Of course, Eq.~(\ref{eq:s0})
is an unequality as an identity: not just as an accident. 
These choices of equations of state (with some simplifications) will be enough as it will be seen.

We have some anisotropy. It has a direction (at the beginning surely the beam
direction), and a measure. So it is a vector. We dub it $t^i$ (anyways, letters
$t\, \&\, u$ are neighbours). Surely $t^i$ is spacelike. So the local state has at least
two preferred vectors: timelike unit $u^i$, and spacelike non-unit $t^i$. The most
general $T^{ik}$ with two preferred vector fields is:
\begin{eqnarray}
T^{ik} &  = & \alpha u^i u^k + \beta(u^it^k + t^i u^k) + \gamma t^i t^k + (d^i u^k + u^i d^k) \nonumber \\
       &    & + (b^i t^k + t^i b^k) + c^{ik}
\label{eq:Tiki}
\end{eqnarray}
where
\begin{equation}
d_r u^r = d_r t^r = b_r u^r = b_r t^r = c_{ir} u^r = c_{ir} t^r = 0
\label{eq:du}
\end{equation}
Vector $d^i$ is well known as heat current; in many cases local equilibrium
approach neglects it as, e. g., the perfect fluid approximation. If so, we may neglect
$b^i$, as well. We may assume that $c^{ik}$ (a 2*2 tensor) is "as isotropic as possible".
Then
\begin{equation}
T^{ik} = e u^i u^k +  \beta (u^i t^k + t^i u^k) + k \{g^{ik} + u^i u^k - t^i t^k / t^2\}
\label{eq:Tik2}
\end{equation}
where we wrote $\alpha \rightarrow e$ due to the usual definition
\begin{equation}
T^{ik} u_i u_k \equiv e
\label{eq:Tiku}
\end{equation}
OK, this seems to be a toy model, and maybe it is too simplified. But now the
energy-momentum tensor is inherently anisotropic, and still we shall get in
due course all the evolution equations from Eqs.~(\ref{eq:T0},\ref{eq:n0})
and from Uneq.~(\ref{eq:s0}); excepting some simple "material equations".

As for the scalar coefficients, equations of state are needed; $\beta$ is 0 in a
mirror-symmetric state, see~\cite{luk902} for the statistical physical approximation;
here we remain at $\beta$=0 for simplicity. The entropy current vector cannot be
anything else than
\begin{equation}
s^i = s u^i + z t^i
\label{eq:si}
\end{equation}
because of the approximation (7).  As for the coefficient $z$ we shall get an equation immediately.

Then Eq.~(\ref{eq:T0}) is an evolution equation for the extensive density $e$, plus the
equation for the acceleration (so for $u^i$). Eq.~(\ref{eq:n0}) is the equation for $n$. We still
need an evolution equation for the anisotropy $t^i$ , but we have Uneq.~(\ref{eq:s0}).

Here a short interlude is needed. Obviously the extent of anisotropy is
characterized by the absolute value of the anisotropy vector; this quantity will
be called $t$, which is now not the time. (Sorry for the convention; it goes
back to 1990.) We may use the Ansatz that the direction of $t^i$ is the beam
direction. However, surely, $t$ is not an extensive; it looks like rather as
a specific extensive. However, a good candidate seems to be e.g.
\begin{equation}
Q = V n t
\label{eq:Qvnt}
\end{equation}
and then the extra extensive density is
\begin{equation}
q \equiv n t
\label{eq:extr}
\end{equation}
and the proper thermodynamic potential density $s$ is
\begin{equation}
s = s(e,n,q)
\label{eq:senq}
\end{equation}
in the simplest case. However, just for now for simplicity let us use
the noncanonical quantity $\hat{s}$
\begin{equation}
\hat{s} \equiv \hat{s}(e,n,t) = s(e,n,q=nt)
\label{eq:hats}
\end{equation}
For a straightforward but tiresome step see Appendix~\ref{app:tra}. The results is a few equations
and unequalities as follows:
\begin{eqnarray}
&& k = p (e,n,t) + \omega T \hat{s},_t + \delta u^r_{;r} \nonumber \\
&& p = T(\hat{s} - n \hat{s},_n - e \hat{s},_e) \nonumber \\
&& z = \beta / T        \nonumber \\
&& \omega \geq  0 \nonumber \\
&& \delta \geq  0 \nonumber \\
&& 1 / T \equiv \hat{s},_e
\label{eq:uneqs}
\end{eqnarray}
where
\begin{equation}
Dt = \lambda + \nu(T_{,r}+T u_{r;s} u^s) t^r + \theta t^r t^s u_{r;s} + \omega u^r_{;r}
\label{eq:Dt}
\end{equation}
and
\begin{eqnarray}
&& \hat{s},_t \lambda \geq  0 \nonumber \\
&& \nu = \beta /T^2 \hat{s},_t \nonumber \\
&& \theta = (k/t^2 - \gamma) /\hat{s},_t
\label{eq:snu}
\end{eqnarray}

Eq.~(\ref{eq:appa}) shows that $\lambda$ contains the characteristic time for the decay
of the longitudinal (=beam direction) momentum, so a weighted cross section.
As we told, just now $\beta=0$. As well, we can take the approximation $\omega=\delta=0$
for the present. And then $\lambda$ is the only new quantity needed.
It governs "thermalisation".

Then we have got the full system of evolution equations as:
\begin{eqnarray}
&&      D n + n u r_{;r} = 0 \nonumber \\
&&      D e + (e + p + q) u^r_{;r} = 0 \nonumber \\
&&      D u + \{D (p + q) + (p + q),_x\} / (e + p + q) = 0 \nonumber \\
&&      D q + (q + q / v) u^r_{;r} - \lambda  = 0
\label{eq:D4diff}
\end{eqnarray}
where x is the beam direction, and for the connection of $q$ and its
canonical conjugate $v$ see \cite{luk89}; the connection, of course, depends on
the equations of state. The system  (Eq.~(\ref{eq:D4diff})) contains the evolution equations
for all independent extensive densities (now for simplicity's sake $e$ is
one of the extensive densities and $s$ is the potential density;
remember Eq.~(\ref{eq:appa})) plus for the velocity; so we can calculate the
local state plus the energy-momentum tensor the next moment.

We are now ready with the evolution equations; still the equation of
state must be specified to establish the extensive-intensive relations,
and also $\lambda$, the coefficient of the decay of longitudinal momentum
must be calculated.

The quantity $v$ is the new entropic intensive canonically conjugate to Q=Vq.
Its actual functional form depends on the equation of state. In~\cite{luk900} and~\cite{luk902}
we used a simple enough model when the matter is dominantly 2 cold
interpenetrating particle currents, still there is a temperature T with a
simple parabolic compression potential and a gluon radiation field. \cite{luk89}
is nicer, with a Skyrme potential in the hadronic phase; but now we are
giving only the formulae for the quark phase, where
\begin{eqnarray}
      p & = & nT + (\pi^2 / 30) T^4 + (K / 9) n_0 x^2 (v \  arctg(v) - 1   \nonumber \\
        &   & + x (2 / y - y)) \nonumber \\
      e & = & m n y + (3/2) n T + (\pi^2 / 10) T^4 + (K / 18) n (x - 1)^2  \nonumber \\
        &   & + (K / 9) n (y - 1) (x / y^2 ((y + 1) (x / y - 1) \nonumber \\
        &   & + (1 - x^2 / y) / 2) \nonumber \\
      q & = & m x n_0 ln(v + y) + (K / 9) n_0 ( (x / 2) ln(v + y)  \nonumber \\
        &   & - x^2 arctg(v) + v x^2 ( (3 x / 2 y) - 1 / y^2 - x v^2 / y^3) ) \nonumber \\
      y & \equiv & (1 + v^2)^{1 / 2} \nonumber \\
      x & \equiv & n / n_0 \nonumber \\
      n_0 & = & 0.16 fm^{-3} \nonumber \\
      m & = & 938 MeV
\label{eq:peqyxnm}
\end{eqnarray}
These equations are somewhat implicit, but hence $v$ can be calculated as function
of $e$, $n$ and $q$, and then the evolution Eqs.~(\ref{eq:D4diff})  are complete indeed.

As for $\lambda$, Eq.~(\ref{eq:D4diff}) shows that it is the decay rate of the density of
extra momentum in beam direction $nt$. In ultrarelativistic situations, as now,
the momentum distribution in beam direction has two peaks with widths $\sim$$T$ and
separation $\sim$$t$$\sim$$E_{beam}$. So in first approximation the distribution is sharp.

Now consider a collision. Most collisions happen between particles moving
oppositely with almost $\pm P_{beam}$. So it is enough to evaluate such a collision.
The details go into Appendix~\ref{app:eva} because of unavoidable inconsistencies in the
notation.

We would need differential cross sections, but that is not easy for inelastic
collisions at such high energies. We chose the compromise to use the total
(elastic+inelastic) cross section with the shape of the elastic one in the
momentum transfer variable. For the latter we used the curve of Islam and al.~\cite{islam} .

\subsection{Some results and outlook}

We solved the dynamic equations in the range of E/A $\sim$ 1 GeV long time ago~\cite{luk903}.
But the equations can be solved in the 100 GeV - 10 TeV
range as easily as below. Ref.~\cite{luk902} (at 1 GeV/A) gave quite rational results
(for 1 GeV/A). The particle density is cca. 2$n_o$ at the center, cca. $n_0$ at
the periphery and evolves in between. At the beginning T is cca. 150 MeV
in the center, dropping towards the periphery; later slightly higher well
inside. The anisotropy is highest somewhere between the center and periphery,
decreasing slowly in time. Indeed, anisotropy exists when both left-moving
and right-moving particles are present in the same volume element; because
of transparency some anisotropy remains even until the end of the overlap
(and afterwards the calculation must stop, with a stage of breakup).

At 200 GeV/nucleon-pair the transparency will be higher. So while the kinetic
energy would be much enough for deconfinement phase transition, it would
be transformed very ineffectively into temperature, and even more ineffectively
into compression. So without full dynamic calculations there is no unequivocal
prediction for or against deconfinement phase transition.

But the full dynamic system of equations is given above and can be used.
And such a scheme is explicit for/in the generation of the entropy.

\section{Numeric results}
\label{chap:numr}

Our simplifications for the entropy production of the collision are as follows:

\begin{enumerate}
\renewcommand{\labelenumi}{\alph{enumi})}
\item Slab geometry (it does not seem too important).
\item Homogeneity (more or less true due to interpenetration).
\item The equations of state given in Section~\ref{chap:form} with K=0.
\item No collision after crossing over (this is more or less true).
\item Breakup just afterwards.
\end{enumerate}

\begin{figure}
\resizebox{0.5\textwidth}{!}{%
\includegraphics{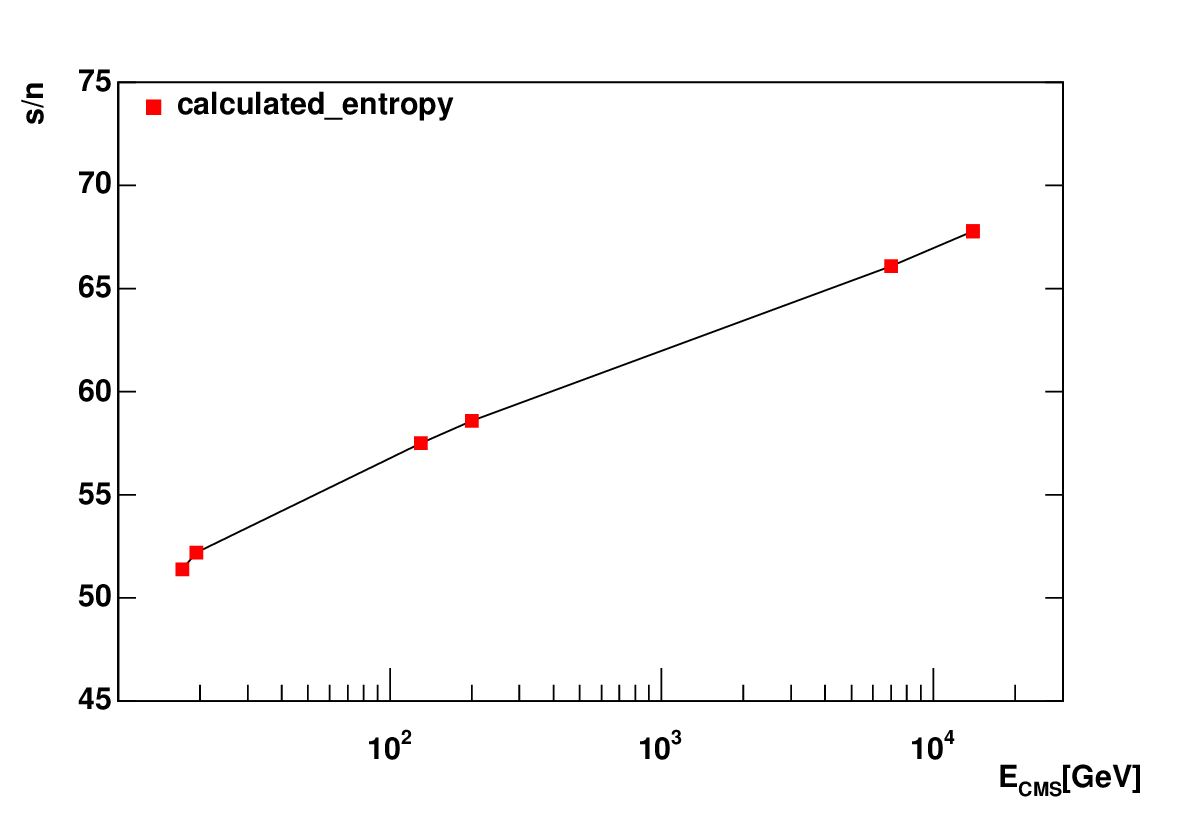}
}
\caption{Calculated specific entropies at different center of mass energies. The solid
         line is drawn for guiding eyes. }
\label{fig_entropies}
\end{figure}

With these simplifications and with Islam's differential cross sections~\cite{islam} 
we have numerically solved Eqs.~(\ref{eq:D4diff}) and we have calculated the S/N 
production for the accelerator energies used 
so far, as well as for LHC. The results are shown in 
Table~\ref{table_entropies} and Fig.~\ref{fig_entropies}.

We have calculated S/N of central Pb+Pb collisions for 6 CM energies 
17.3 GeV, 19.4 GeV (158 \& 200 GeV fixed target), 
130 GeV, 200 GeV, 7000 GeV \& 14000 GeV, respectively. 
While the last 2 energies seem to be 
possible only for p+p collisions, the calculations are meaningful, as it will be seen. 
The results are shown by Fig. 1.

The analysis of the curve shows that a power fit is not too bad but with a hardly 
interpretable power law 0.03, while a logarithmic fit S/N $\approx$ 45.31 + 2.38ln($E_{CM}$ [GeV]) 
is even better. The logarithmic behaviour is not surprising in an ultrarelativistic 
regime without energy scale.

We cannot determine the error of the calculation, but we can give mean $\chi^2$ 
deviations from the logarithmic curve. However, doing this one can see that the 
main deviation would come from the two SPS data points; obviously below 20 GeV we are 
not in the true ultrarelativistic regime. Removing these points the fit is 
much better, resulting in  S/N $\approx$ 47.019 + 2.168 ln($E_{CM}$ [GeV]), with a $\chi^2$
error $\delta$(S/N) = 0.14. The error is quite low, and we will use it from 200 GeV onward. 
So, for the relevant energies the specific entropies are given in Table 1. 
The dynamic calculations give S/N=66.1 for 7000 GeV and 67.8 for 14000 GeV.
\begin{table}[]
\begin{center}
\begin{tabular}{|l|l|}
\hline
ECM [GeV]    &S/N    \\
\hline
17.3         &51.4   \\
\hline
19.4         &52.2   \\
\hline
130          &57.5   \\
\hline
200          &58.6   \\
\hline
2760         &64.2   \\
\hline
5520         &65.7   \\
\hline
\end{tabular}
\end{center}
\caption
{
Specific entropies at various energies of center of mass.
}
\label{table_entropies}
\end{table}

\hyphenation{Letes-sier}

The entropies are in the expected range. Indeed, Letessier, Rafelski \& Tounsi
read out almost these values from yields and energy distributions cca. 15 years
ago. At $E_{beam}$ / A = 200 GeV, which is the second energy of the Table,
$\sqrt{s}$ = 19.4 GeV. In 1993 the value was $50\pm 4$ ~\cite{let93}, in 1995
42 $\langle$ S/N $\langle$ 48 ~\cite{let95} and in 1996 35 $\langle$ S/N $\langle$ 60 ~\cite{let96},
any of them an excellent first guess.

\section{Application of rehadronization schemes}
\label{chap:app}

The most direct experimental facts about QCD are the particle yields
at the detectors. However, these particles are hadrons, coming from a
hadronic phase after rehadronisation. Only their quark constituents
keep the memory of the QGP phase. Therefore, the actual rehadronisation
process seriously influences the yields, but we do not know a priori the
correct rehadronisation model.

Here we use a variety of rehadronisation models simultaneously: each one
would lead to a set of yields, and at the end one of them or at least
a combination of them might be satisfactorily near to measured yields
at lower energies. Then we may take this model or the combination or an
extrapolation. This decision
will be done in the next Section; now let us see the models used here.
Calculations in models 1-10 were made a few times; of course, in
contexts of much lower energies \cite{luk91,zim93,arv91,kfki93,luk96}.
We consider here 10 rehadronisation models (+1 without QGP)
from amongst the virtual infinity of them.

Models 1-8 assume a QG plasma at maximal compression; in the QG plasma
there is complete thermal, chemical and mechanical equilibrium
(processes are rapid enough). Then an expansion starts and somewhere
the matter reaches the phase boundary.  As a matter of simplification
the phase boundary is fixed on the p-T plane, and for the actual choice
see~\cite{koc86}. Then rehadronisation starts, but the growing HM droplets remain
in equilibrium with the QG environment (no supercooling \&c.). The process
goes until the last remnants of QGP vanish: then we detect the hadrons.

Even this scenario means an infinity of models, of which here we take 8,
as follows. We have 3 properties of the models, and $2^3=8$.

1) Gluon fragmentation may occur during rehadronisation: we may neglect it
(0) or may calculate it as 15\% $s\overline{s}$ and 85\% $q\overline{q}$.
For the suggestion of this particular ratio see~\cite{koc86}; for most probable
strength see~\cite{zim93,kfki93,koc86}.

2) The probability of hadronic combinations may come from the quark numbers
by simple combinatorics (0) or they may contain an extra factor $\sim e^{-F/T}$ (1),
where F is the free energy. This latter idea is obtained from the numbers
of final state microstates $\sim e^{S}$  \cite{kam86}, and for technical reasons here we will
approximate F by the hadronic rest mass $m$.

3) The hadronic phase may have a compressibility of $240 MeV/fm^3$ (1) or not (0).

The 3 numbers are regarded digital, and then add 1: so they are indeed
Models 1-8. Model 9 is the absence of QGP phase at all, for simplicity's sake with
complete thermal, chemical and mechanical equilibrium throughout
the whole collision, and Model 10 is a sequential fission one~\cite{arv91}.
A comparison of the results of Models 1-10 was given in~\cite{holba} up to SPS energies.
We will add Model 11 in due course.
All of these hadronisation models end in stables. Models ending in resonances are not
considered here, because very various such models would be possible. We will later
discuss the consequances of this approximation.

12 hadrons are included into the scheme, with the assumed QG precursors, as indicated:
\begin{eqnarray}
&&      N      = qqq \nonumber \\
&&      Y     (\equiv \Lambda \, \& \, \Sigma) = qqs \nonumber \\
&&      \Xi    = qss \nonumber \\
&&      \Omega = sss \nonumber \\
&&      + antibaryons + \nonumber \\
&&      \pi    = q\overline{q} \nonumber \\
&&      K      = q\overline{s} \nonumber \\
&&\overline{K} = s\overline{q} \nonumber \\
&&      \eta   = s\overline{s}
\label{eq:hadrons}
\end{eqnarray}
for a while without distinguishing $u$ and $d$.
All of these are "stables", i.e., they do not decay in strong interaction.
Obviously in the initial state there
is a slight $d$ excess, which may cause a bias (see next Section), but
not too much, and the last mesonic representation is highly debatable
but $\eta$ will not in fact be used explicitly.

\begin{figure}
\resizebox{0.5\textwidth}{!}{%
\includegraphics{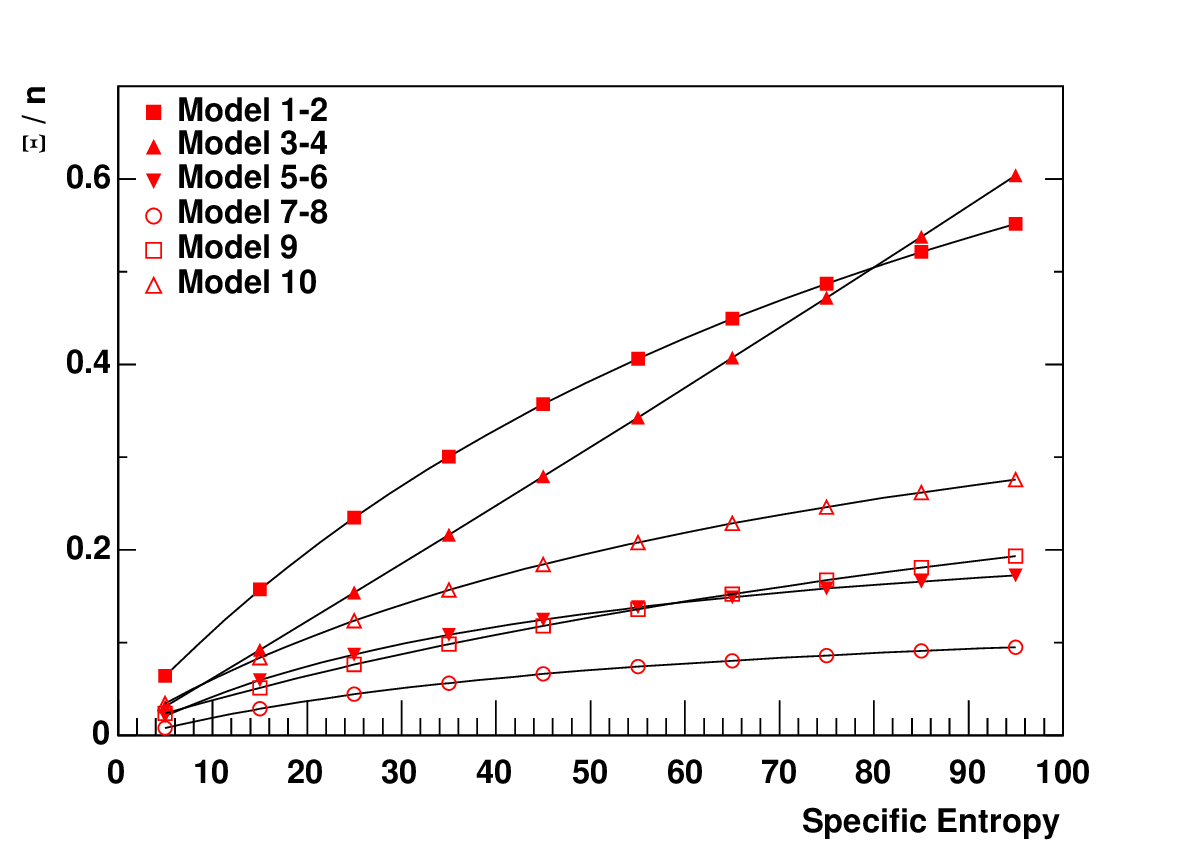}
}
\caption{\small $\Xi / n$ ratios of model predictions above S/N=5. The solid lines
are drawn for guiding eyes.}
\label{fig-label1}
\end{figure}

\begin{figure}
\resizebox{0.5\textwidth}{!}{%
\includegraphics{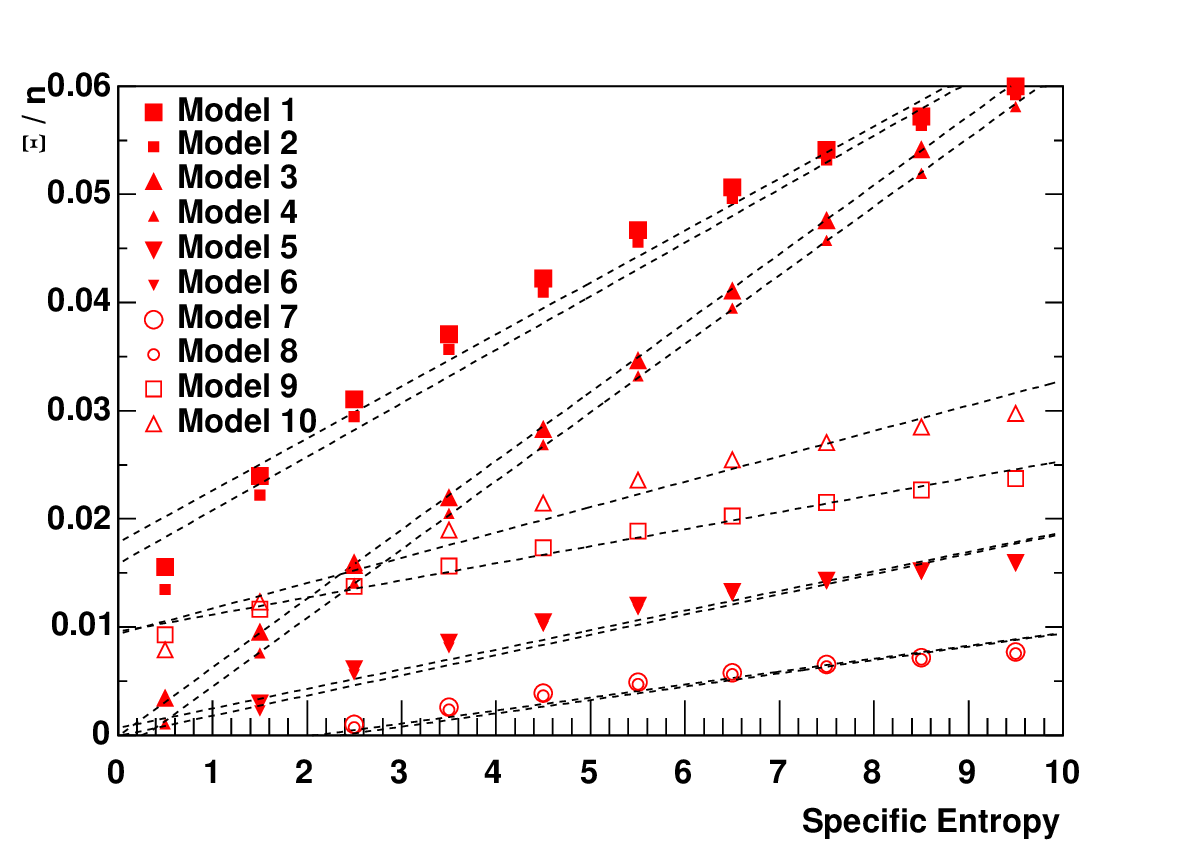}
}
\caption{\small $\Xi / n$ ratios of model predictions between S/N=0 and 10. 
The solid lines are drawn for guiding eyes.}
\label{fig-label2}
\end{figure}

Then we  get unique predictions for the 12 yields if the model and the
specific entropy are fixed. To demonstrate the model and specific entropy
dependences, consider the ratio $\Xi / n$, for the Models 1-10 between
S/N = 5 and 95 and separately between 0 and 10 
on Fig.~\ref{fig-label1} and Fig.~\ref{fig-label2}, respectively.

As it can be seen, for Models 1-8 there are 8 different
predictions  at low S/N's, but model pairs practically coincide
for S/N$>$20. So henceforth, it will be superfluous at RHIC and LHC energies to refer
separately to Models 2,4,6 \& 8.

Of course, very probably none of the rehadronisation models will be
satisfactory because i) the rehadronisation processes do not clearly separate;
and ii) some of the present models have fixed parameters which are
not necessarily the true ones.
Problem i) is seldom formulated explicitely, although it is serious.
As an example, it is quite possible that in rehadronisation the result
is not $p=uud$ but $\Delta^+=uud$ with parallel spins. Then $\Delta^+$ decays
in the hadronic phase giving extra mesons, mainly pions. None of the the
11 rehadronisation models include resonances, for obvious difficulties.
One of them is that very probably some $uud$ triads would correct the spin positions
still in quark phase, while some $\Delta^+$s are formed and then decay in the
hadronic phase; and we do not know, how much. (There is some indication for such
decays and extra $\pi$'s at 200 AGeV RHIC as it will be seen in Section~\ref{chap:app})
We think only $\pi$ numbers will be sensitive on this.
As for to overcome the second problem we can try
with linear combinations of models with weights fitted to the
observed yields. However, at 200 GeV the available yield ratios are limited.
So we give here the method only for 3 models selected in the same time.
Let the models be denoted by f, g and h, respectively; the index $i$
denotes the actual ratio. Then the actual yields are given by the combinations

\begin{eqnarray}
          R_i = a(E)f_i + b(E)g_i + (1-a-b)h_i + \sigma_i
\label{eq:Ri}
\end{eqnarray}
where the $\sigma_i$'s are the errors of the combined model, to be minimized in average.
This means that
\begin{eqnarray}
          (N-2)\sigma^2 \equiv \Sigma_{i=1}^N \sigma_i^2 = min.
\label{eq:sigma2}
\end{eqnarray}
Hence
\begin{eqnarray}
&&	a\Sigma_{i=1}^N (f_i-h_i)^2        + b\Sigma_{i=1}^N (f_i-h_i)(g_i-h_i)  \nonumber \\
&&	= \Sigma_{i=1}^N (R_i-h_i)(f_i-h_i)                                      \nonumber \\
&&	a\Sigma_{i=1}^N (f_i-h_i)(g_i-h_i) + b\Sigma_{i=1}^N (g_i-h_i)^2         \nonumber \\        
&&	= \Sigma_{i=1}^N (R_i-h_i)(g_i-h_i)
\label{eq:fgh}
\end{eqnarray}

Hence we get the best weights, and substitute them back Eq.~(\ref{eq:fgh}) gives the average $\sigma^2$.
\begin{table}[h]
\begin{center}
\begin{tabular}{|l|l|l|l|l|}
\hline
Model        &158 GeV     &200 GeV       &130 GeV   &200 GeV    \\
             &SPS         &SPS           &RHIC      &RHIC       \\
\hline
1 \& 2       &0.394       &0.759         &0.012     &0.0503     \\
\hline
3 \& 4       &1.326       &2.819         &0.009     &0.0306     \\
\hline
5 \& 6       &0.120       &0.141         &0.005     &0.0330     \\
\hline
7 \& 8       &0.165       &0.283         &0.005     &0.0308     \\
\hline
9            &0.020       &0.028         &0.072     &0.0950     \\
\hline
10           &0.084       &0.176         &0.054     &0.1288     \\
\hline
\end{tabular}
\end{center}
\caption
{
The average predictive squared errors of the individual models.
}
\label{table_errors}
\end{table}

Now we are going to compare the measured particle ratios at 158 \& 200 GeV
fixed target (SPS) and 130 \& 200 GeV storage ring (RHIC) experiments to the predictions
of the rehadronisation models 1-8, 10 plus Model 9 without QGP
 to find the best one or the best combination.
We took the ratios from \cite{let96,let93,let95,let97,let972,red02,red022,tom03,bic04,let973,bar90}; 
when we found more than one ratio
but still near to each other, at a few $\sigma$, we averaged with least squares.
(When they were too different we discarded one of them.)
We tried to use low impact factor collisions,
as far as possible. Later Table~\ref{table_ratios} will give the ratios we used in the analysis.
The deviation $\sigma$  is the quantity in Eq.~(\ref{eq:sigma2}): the average "error" of the
rehadronisation model. It is dimensionless. Table~\ref{table_errors} gives the average
squared errors for individual models. However, note that if a single model
is evaluated the divisor changes from N-2 to N-1.

158 GeV fixed target means 17.3 GeV in storage ring situation, at least
for symmetric situations. For 200 GeV the equivalent energy is 19.4 GeV.
As shown by Fig. 2 at the entropies relevant now the first 8 models
give only 4 different results up to 3 digits.

The picture is clear enough: at the two lowermost energies the best
predictions are given by Model 9, so a model with no QGP at all.
Even Model 10, a sequential fission, is much worse.
The higher error at the higher energy probably comes simply from the fact that
we used less ratios there; still Model 9 wins clearly.
So it seems that at
those energies no well developed QGP was formed. This seems to be conform
with the opinion of Ref.~\cite{let96} and~\cite{let973} from 1996-7 even if in the
previous years still the hopes were high for 19.4 GeV, and even if Ref.~\cite{let99}
in 1999 still was optimistic.
Just oppositely, for 130 \& 200 GeVs Ref.~\cite{let06} expects QGP in 2006, and indeed at
this energies QGP models win.  At 130 GeV the best models are 5 through 8;
at 200 GeV even 3 \& 4 are as good, but at the highest energy we
could use only 5 ratios. So we are giving greater weight to the
130 GeV errors; but clearly we have to try with the linear combinations, as well.

There is no doubt at the two lower energies. The three-component combinations
of smaller $\sigma_i^2$ always contain Models 9 or 10; if both then the weight of
10 is either smaller or even negative. With a dual combination of 9 \& 10
the error minimum is at combinations with weight slightly above 1 for Model 9
and slightly negative for Model 10. Since weights either $\rangle$ 1 or $\langle$
0 are aphysical,
this means that in the whole 9-dimensional space of the linear combinations
of Models 1-10 the physical optimum is at (or very near to) the pure Model 9.

For RHIC at 130 GeV the situation is more complicated. Models 5-6 and 7-8 give very
similar small errors. Trial combinations with both 5 and 7 give various
minima but always at aphysical weights. Dual combination of 5 and 7 is optimal
at positive weight of 7 and negative one for 5. Finally an analysis of 5-7
combinations gives monotonous decrease in the physical domain from
pure 5 to pure 7, but the whole difference is small: for Model 5 $\sigma_i^2$=0.0061,
for Model 7 0.0058.

At 200 GeV the result is similar, although the overall error is higher (less
ratios were used) and Model 3 is very sligthly better then Model 7.
The squared average is generally high (which is the reason that we remain at Model 7 as best).
We refer to this point in due course.

All the models 5-8 contain gluon fragmentation. So at least from the particle yields,
it seems as if both a well-developed QGP and substantial gluon fragmentation
appeared somewhere in the range between 20 and 130 GeV. It would be interesting
to see yields between 20 and 130 GeV; but for now we go directly to 5520 GeV.
We cannot expect Models 9 \& 10 to overcome again at high entropy; our best
educated guess is Model 7(-8). However, we give predictions of Model 5, as well,
having been almost as good at 130 GeV. Because of the substantial presence of
gluon fragmentation we give a Model 11, as well, which is Model 7 but with
exaggerated gluon fragmentation, tuned up to the energetically possible limit.
(In our language it is 1.36 instead of 1 for the gluon fragmentation
parameter~\cite{zim93,kfki93}.) Also, Model 11 may mimic somewhat
rehadronisation into resonances (both processes resulting in extra light mesons).

Now, we are in the position to give the 5.52 TeV predictions.
(This energy is valid only for Pb+Pb, being then Z=82 and for average isotopic 
coposition A=207. However this energy is almost constant in a wide range, 
e.g. in Au+Au it is 5.61 TeV, for example.)
First we give Table~\ref{table_quant} for stables defined in Eq.~(\ref{eq:hadrons}),
still not distinguishing between the light quarks.
Then we list 4 simplifications inherent in the calculations but easy to
estimate for the errors caused, and even correct it in some cases.

\begin{table*}
\begin{center}
\begin{tabular}{|l|l|l|l|l|l|l|}
\hline
Particle           &Model 5    &\bf Model 7   &Model 9      &Model 11       &Model 7        &Mass      \\
                   &           &              &             &               &(with S/N=75)  &(MeV)     \\
\hline
N                  &528.79     &\bf 475.95    &389.19       &585.18         &512.60         &939       \\
\hline
$\overline{N}$     &326.50     &\bf 301.58    &152.70       &365.80         &350.83         &939       \\
\hline
Y                  &610.63     &\bf 711.38    &261.25       &658.50         &801.15         &1174      \\
\hline
$\overline{Y}$     &412.91     &\bf 479.61    &113.76       &461.67         &557.46         &1174      \\
\hline
$\Xi$              &79.09      &\bf 38.49     &59.65        &26.79          &44.11          &1318      \\
\hline
$\overline{\Xi}$   &64.00      &\bf 30.98     &32.14        &22.33          &35.94          &1318      \\
\hline
$\Omega$           &19.79      &\bf 1.01      &8.03         &0.35           &1.17           &1672      \\
\hline
$\overline{\Omega}$&19.79      &\bf 1.01      &5.53         &0.35           &1.17           &1672      \\
\hline
$\pi$              &3755.23    &\bf 3686.70   &1775.05      &4104.40        &4212.90        &138       \\
\hline
K                  &1285.38    &\bf 1388.59   &965.63       &1273.83        &1579.55        &496       \\
\hline
$\overline{K}$     &1057.50    &\bf 1141.80   &757.30       &1068.06        &1319.53        &496       \\
\hline
$\eta$             &162.89     &\bf 125.60    &358.21       &96.23          &144.64         &549       \\
\hline
\end{tabular}
\end{center}
\caption
{
The quantities of Point 4 at 5.52 TeV. Y stands for ($\Lambda,\Sigma$).
The bold column is our favoured prediction.
}
\label{table_quant}
\end{table*}

{\it Point 1. Slight Simplification: Insensivity on Charge and its Consequences}

So far, we have not distinguished between the 2 light quarks. Therefore, N($p$)=N($n$) \&c.
This is definitely not true for the initial state with heavy nuclei,
although it is still not an absurd approximation.
The final results can somewhat be corrected according
to the u/d ratio of the initial condition; but then the ratios should
be calculated individually for different beams. For Pb+Pb N($u$)/N($d$)=0.87.
Here we are going to make the correction.

In the charge insensitive approximation:
\begin{eqnarray}
&&      \textnormal{If} \,\, N(u)=N(d), \textnormal{then} \nonumber \\
&&      N(p)=N(n) \nonumber \\
&&      N(\Sigma^+)=N(\Sigma^-)=N(\Sigma^0)=N(\Lambda)/2 \nonumber \\
&&      N(\Xi^0)=N(\Xi^-) \nonumber \\
&&      \textnormal{ and similarly for antibaryons,} \nonumber \\
&&      N(\pi^+)=N(\pi^-)=N(\pi^0)/2 \nonumber \\
&&      N(K^0)=N(K^-)   \nonumber \\
&&      \textnormal{and similarly for the antiparticles.}
\label{eq:NN}
\end{eqnarray}
(For the Y's see App. C.)

Now, consider a well-developed QGP.
Really, N($d$) $>$ N($u$).  However, the asymmetry is
smaller then in the initial state, because the processes creating
quarks \& antiquarks are charge symmetric. At the state just before the
rehadronisation
\begin{eqnarray}
&&      N(d)+N(u)=1242+N(\overline{d})+N(\overline{u}) \nonumber \\
&&      N(s) = N(\overline{s})
\label{eq:Nq}
\end{eqnarray}

(The second formula is invalid for Model 10, and there are no quarks
at all for Model 9.)

All antiquarks have been produced in charge-symmetric processes and all
quarks above the initial 1242 ones, as well. Since the rehadronisation uses
the same pool for any hadron, the final result is (with $R \equiv d/u$) cca.
as follows. Left hand sides are the charge asymmetric yields, right hand sides
are the quantities in Eq.~(\ref{eq:hadrons}).
\begin{eqnarray}
&&      p = N/(R+1),              \,\,\, \overline{p} = \overline{N}/2       \nonumber \\
&&      n = N*R/(R+1),            \,\,\, \overline{n} = \overline{N}/2       \nonumber \\
&&      \Lambda =Y*   /2         ,\,\,\, \overline{\Lambda} = \overline{Y}/2 \nonumber \\
&&      \Sigma^-=Y*R^2/2(R+1)^2  ,\,\,\, \overline{\Sigma^-}= \overline{Y}/8 \nonumber \\
&&      \Sigma^0=Y*  R/(R+1)^2   ,\,\,\, \overline{\Sigma^0}= \overline{Y}/4 \nonumber \\
&&      \Sigma^+=Y*  1/2(R+1)^2  ,\,\,\, \overline{\Sigma^+}= \overline{Y}/8 \nonumber \\
&&      \Xi^0   =X*  1/(R+1)     ,\,\,\, \overline{\Xi^0}   = \overline{X}/2 \nonumber \\
&&      \Xi^-   =X*  R/(R+1)     ,\,\,\, \overline{\Xi^-}   = \overline{X}/2 \nonumber \\
&&      \Omega^-=\Omega          ,\,\,\, \overline{\Omega^-}= \overline{\Omega}       \nonumber \\
&&      \pi^+   =\pi*(1/2)/(R+1) ,\,\,\, \pi^0 = \pi/2, \,\,\, \pi^-         \nonumber \\
&&	=\pi*(1/2)*R/(R+1) \nonumber \\
&&      K^+     =K      *1/(R+1) ,\,\,\, \overline{K^+}= \overline{K}/2      \nonumber \\
&&      K^0     =K      *R/(R+1) ,\,\,\, \overline{K^0}= \overline{K}/2      \nonumber \\
&&      \eta    =\eta
\label{eq:yields}
\end{eqnarray}

For colliding nuclei others than Pb the charge-symmetric Table~\ref{table_quant}
scales with $N_B$, the initial $u+d=3N_B$, and R can be calculated from the initial u/d
and from the rescaled Table~\ref{table_quant}. For Au+Au N($u$)/N($d$) = 0.88.

{\it Point 2. A Minor Neglection in the ($\Sigma,\Lambda$) Segment}

There is a slight difference between the $\Sigma$  and $\Lambda$  masses.
This here will be ignored (as it will be seen, $\Sigma^0$ is not really
observable, anyway).  There is an even slighter difference amongst
the $\Xi$  masses, also ignored.
The reason of the mass differences is among $\Sigma$s and between $\Xi$'s that the quarks are the final
components. So minimal $\Sigma$ energy is in uus combination. These differences are a few
MeV's. The mass difference between $\Lambda$ and $\Sigma^0$ is, however, 77 MeV.
While this may still be not too serious, the real problem is that both have uds composition
the relative spin positions differ. Therefore, in the ($\Lambda,\Sigma$) sector the resolution
depends in somewhat arbitrary assumptions until resonances will be included.
Our present assumption is discussed in Appendix~\ref{app:the}; it gives the numbers in Eq.~(\ref{eq:yields})
and this problem will also appear in next Section.

{\it Point 3. The Stables}

Our model includes the "stable hadrons" of the particle physicist convention,
that is those which would be stable were the s quark stable (no weak interaction).
These "stables" are:
\begin{eqnarray}
&&      n, p, \Sigma^+, \Sigma^0, \Sigma^-, \Lambda, \Xi^0, \Xi^-, \Omega^- \nonumber \\
&&      \textnormal{the respective antihyperons, and the mesons} \nonumber \\
&&      \pi^+, \pi^0, \pi^-, K^-, K^0_L, K^0_S, K^+, \eta.
\label{eq:stables}
\end{eqnarray}

They do not decay on $10^{-23}$s scale. Without distinguishing between $u$ an $d$
(charge insensitive case) there were 12 stables as in Eq.~(\ref{eq:hadrons}); distinguishing
them there will be 26 ones as in Eq.~(\ref{eq:yields}).  However,
really $\eta$ is rather obscure and it contains both $q\overline{q}$ and $s\overline{s}$.
Now, observe that $\Sigma^0$, $\pi^0$ and $\eta$ are not detected. (They are "stables", but
they do not reach the detectors). The lifetimes are
long enough to survive breakup but not long enough to directly detect them.
So it is only of secondary importance what is the exact quark content of $\pi^0$ \& $\eta$.
(Really, in the $O^-$ nonett two heavy and not too familiar measons, $\eta$ and $\eta$', are
represented by our symbolic $\eta$.)
Here we regard them as the lighter composed of a light quark pair and the heavier
of a strange pair.
In addition obviously, Models 1-8 contain a very important neglection, mentioned already briefly.
Namely, some hadrons
probably will be first resonances not stables. But both rehadronisation and resonance
decay are in $10^{-23}$s characteristic time, so some resonances decay in the hadronic
matter, giving extra light mesons, mostly pions. This process is not included in
the models, and it would be rather difficult to do it. Here, we only declare that
the numbers are very probably underestimated, therefore, in all the models.

Also, there is the well known $K^0$ problem. For all this, see the next Point 4.

{\it Point 4. The Observables}

For handling $\Sigma^0$, $\pi^0$ \& $\eta$ observe that neglecting leptons \&
photons $\Sigma^0$ decays solely to $\Lambda$, $\pi^0$ decays solely to nothing,
and $\eta$ gives a $\pi^0$ (which then decays into nothing) or either nothing else or
a $\pi^+\pi^-$ pair. Since $\Sigma^0$, $\pi^0$ \& $\eta$ decay before
the detectors, in them we get:
\begin{eqnarray}
&&      \Sigma^0 \rightarrow 0                    \nonumber \\
&&      \Lambda  \rightarrow \Sigma^0 + \Lambda   \nonumber \\
&&      \pi^0    \rightarrow 0                    \nonumber \\
&&      \eta     \rightarrow 0                    \nonumber \\
&&      \pi^+    \rightarrow \pi^+ + 0.29 * \eta  \nonumber \\
&&      \pi^-    \rightarrow \pi^- + 0.29 * \eta
\label{eq:observ}
\end{eqnarray}
where  $\rightarrow$ does not mean any temporal sequence but a calculation of detectables.
For $K^0$'s the detectable components are $K^0_L$ and $K^0_S$, for which
\begin{eqnarray}
&&      K^0_L    =K^0_S=(K^0+\overline{K^0})/2
\label{eq:observK}
\end{eqnarray}
This way we still have, say, $\Sigma^0$ if it can be reconstructed, which is sometimes claimed.

\section{Discussion}
\label{chap:dis}

Earlier we have told that at lower energies (158 \& 200 GeV/nucleon
fixed target) the model best for reproducing the yield ratios amongst
the 11 ones included in the present survey was Model 9, the only
one without QGP phase at all. This is a signal that at these energies
pure quark phase is still absent. However, at 130 \& 200 GeV per nucleon
storage ring energies the two best models are 5 ("100") and 7 ("110").
Model 7 is marginally better, and linear combination does not
improve the performance: at 200 GeV the situation is not clear but
Model 7 is better then Model 5.
Note that individual yields may differ between
Models 5 \& 7; only they are near to each other in overall performance.

Since this fact indicates well-developed QGP already at 200 GeV, we do
not expect anything else at 5.52 TeV. Therefore, our favourite prediction
is Model 7; but we do not deny a slight chance from Model 5. As for Model 9,
the one without QGP, we give its yields only for comparison.
Model 11 is Model 7 but with the maximal possible gluon fragmentation. Finally,
we give also the predictions of Model 7 for S/N = 75 (instead of the expected 65.7);
the reason will come.
The final results are in Table~\ref{table_observ}; then you can calculate
any ratios of the observables taking into account Eqs.~(\ref{eq:observ}, \ref{eq:observK}).
As for the error of the prediction, in theory the $\chi^2$
test at 200 GeV suggests ~$\pm 0.02$ in ratios,
but the $\sigma$'s in Table~\ref{table_errors} show that there are serious
biases, too.

Of course most biases/systematic errors are unknown,
but we can estimate 2 of them.

First, as told in Point 1 of Section~\ref{chap:app}, the simplified model overestimates the number
of quarks $u$. Because of heavy nuclei in the beam the true $u/d$ ratio is
$\approx$  0.87 ($Pb+Pb$). Now, all in the ratios $\Sigma^+/p$, $\Omega^-/\Xi^-$
and $\pi^-/K^-$ the ratio N($s$)/N($d$) appears, and N($d$) is in reality
greater than in the model by  cca. 6 \% relative. However, note that
this distortion is less in Model 7 where the gluon fragmentation somewhat
dilute the $d$ surplus, and even less in Model 11. And in Table~\ref{table_observ}
we will give the charge-sensitive yields according to Eq.~(\ref{eq:yields}).

The second type of error may come from the thermodynamic nature of the models;
in them particle number conservations are not explicit. But the final
particle numbers can check the accuracies.
Table~\ref{table_checks} summarizes some results of Table~\ref{table_quant}.

\begin{table*}
\begin{center}
\begin{tabular}{|l|l|l|l|l|l|}
\hline
Total $N^o$     &Initial      &Model 5 &\bf Model 7  &Model 9 &Model 11    \\
\hline
Particle        & -           &8322.50 &\bf 8382.70  &4878.44 &8653.47     \\
\hline
Strangeness     & 0           &0.02    &\bf 0.00     &1.68    &-0.01       \\
\hline
Baryon          & 414         &415.10  &\bf 413.65   &413.99  &420.68      \\
\hline
\end{tabular}
\end{center}
\caption
{
Conservation checks for Pb+Pb collisions with 5.52 TeV/nucleon-pair.
}
\label{table_checks}
\end{table*}
We can see that Models 7 \& 9 are excellently consistent with the conservation
laws, and the relative errors are low even for Models 5 \& 11, because the baryon
number \& strangeness conservations should be maintained in the presence of
lots of particles with negative quantum numbers. If we relate the above errors
to the averages of numbers of particles with + \& - quantum numbers, then at the
worst cases the relative error is 0.1\% for strangeness and 0.7\% for baryon number.

Observe the tremendous particle production in the collisions. Just after breakup
there are at least an order of magnitude more particles present than in the initial
condition. As for the differences between the total yields of the models we see
some 1\% difference between Models 5 \& 7, hardly more between 7 \& 11 and the
only serious difference is Model 9 without QGP phase.

Charge conservation will be checked after the charge-asymmetric Table~\ref{table_observ}.

Table~\ref{table_Etot}  gives us the data for estimating the efficiency turning the beam
energies into "something". At the initial condition there are 414 nucleons, each with
2760 GeV kinetic energy. Just after rehadronisation/breakup some of the kinetic
energy has turned into rest mass, some into internal energy in thermodynamic sense,
and some energy has gone away in the beams. Calculating with a 160 MeV temperature
at rehadronisation as approximately got in Models 1-8 \& 11, the numbers are as follows:
where $E_{transf}$ does include the rest masses, but not the ones present in the beginning.
At the initial condition $E_{transf}$ was 0, and $M_{tot}$/414 was 0.94 GeV.
\begin{table*}
\begin{center}
\begin{tabular}{|l|l|l|l|}
\hline
Model \# \ \ \ \ & $E_{tot}/414$ [GeV]   & $M_{tot}/414$ [GeV]  & $E_{transf}/414$ [GeV] \\
\hline
    5        & 2760                &  9.73              & 13.62               \\
\hline
\bf 7    & \bf 2760            &  \bf 9.80          & \bf 13.72               \\
\hline
    9        & 2760                &  5.77              &  7.66               \\
\hline
   11        & 2760                &  9.78              & 13.86               \\
\hline
\end{tabular}
\end{center}
\caption
{
Conversion of initial kinetic energy $E_{tot}$ into energy transferred to new particles.
}
\label{table_Etot}
\end{table*}

So we can see that the efficiency of conversion is predicted to be
$(3~-~5)*10^{-3}$ at 5.52 TeV. The reason is of course the forward-peaked
differential cross section. (As told earlier if rehadronisation goes
through resonances, not included in the models,
then the resonances yield extra light mesons. This, of course increases
the kinetic energy transferred to masses.
However, the effect is not big. Doubling the number of pions the
last column of Table~\ref{table_Etot} goes up with cca. 1.5)

At last, we are going to investigate the unavoidable consequences of the
present lack of observations at energies between Tevatron and LHC.
In Section~\ref{chap:numr} we evaluated the expected LHC specific entropy and we got 65.7.
However, the experimental differential cross sections are not yet measured
above Tevatron momenta; we have used extrapolations~\cite{islam} assuming that
nothing new happens there. This is probably so; but if we cross the energy
range of "something new" between Tevatron and LHC energies, then in first
approximation a "reasonance peak" is expected, so increasing the cross section
and so the S/N. This "something new" might
be anything (e.g. crossing the supersymmetry energy scale) and we have no preferred
candidate, at all. It will not be necessary either; we can simply
perform calculations with our best, Model 7, with S/N = 75 (some 10\% increase)
for the observables according to Point 5.

Now, we make the charge-sensitive evolution of yields for Pb+Pb. Accepting the
charge-insensitive yields of Table~\ref{table_quant}, we operate as told
earlier for u's and d's in Point 1, so finally:
\begin{eqnarray}
&&      N(u) = 578 + (1/2) [N(q) - 1242] \nonumber \\
&&      N(d) = 664 + (1/2) [N(q) - 1242] \nonumber \\
&&      N(s) = N(\overline{s})         \nonumber \\
&&      N(\overline{u}) = N(\overline{d}) = (1/2) [N(q) - 1242]
\label{eq:Nud}
\end{eqnarray}
with this we get Table~\ref{table_observ}, with stables, with our definite prediction in
column 3, Model 7, S/N=65.7; if any of the other columns will have been proven
superior, one can at least guess hence the reason.
For the detected ones one can use Eqs.~(\ref{eq:observ}, \ref{eq:observK}).
We have chosen this way, because
indirect methods may reconstruct at least $\Sigma^0$ and $\pi^0$.
For comparision, the last column gives figures of Model 7 for 200AGeV at RHIC.
At this point we can check charge conservation. Interestingly,
charge conservation depends on the handling of the ($\lambda,\Sigma$) sector.
We tried two different decompositions of $Y$'s into $\lambda$ and $\Sigma$ which
will be discussed in Appendix~\ref{app:the}.
Exact charge conservation would mean Z=164; at the first choise $Z \approx 174$
for all the models, in the second one $Z \approx 172$. So we used the second for yields,
but our calculations seem to indicate that further supression of $\Sigma$s
do not help in charge conservation. Note, that our rehadronisation models are
thermodynamic, charge asymmetric states are not of minimal energy. Therefore,
thermodynamic models have the tendency to violate charge conservation.
Anyway, the net error 5\% correspond to error/all charged $\approx$ 0.2\%, in the order
of the errors for strangeness conservation.

\begin{table*}
\begin{center}
\begin{tabular}{|l|l|l|l|l|l|l|}
\hline
Particle              &Model 5    &\bf Model 7   &Model 9      &Model 11      &Model 7         &Model 7     \\
                      &           &              &             &              &(with S/N=75)  &(200 GeV RHIC)\\
\hline
p                     &261.53     &\bf 235.41    &190.87       &289.62        &253.85          &221.30      \\
\hline
$\overline{p}$        &163.25     &\bf 150.79    &76.35        &182.90        &175.42          &131.99      \\
\hline
n                     &267.26     &\bf 240.54    &198.32       &295.56        &258.76          &226.66      \\
\hline
$\overline{n}$        &163.25     &\bf 150.79    &76.35        &182.90        &175.42          &131.99      \\
\hline
$\Sigma^+$            &74.68      &\bf 87.01     &31.42        &80.65         &98.23           &78.45       \\
\hline
$\overline{\Sigma^+}$ &51.61      &\bf 59.95     &14.22        &57.71         &69.68           &52.52       \\
\hline
$\Sigma^0$            &152.64     &\bf 177.82    &65.29        &164.61        &200.27          &160.69      \\
\hline
$\overline{\Sigma^0}$ &103.23     &\bf 119.90    &28.44        &115.42        &139.37          &105.04      \\
\hline
$\Sigma^-$            &77.99      &\bf 90.85     &33.92        &83.99         &102.07          &82.29       \\
\hline
$\overline{\Sigma^-}$ &51.61      &\bf 59.95     &14.22        &57.71         &69.68           &52.52       \\
\hline
$\Lambda$             &457.95     &\bf 533.51    &195.91       &493.86        &600.84          &482.12      \\
\hline
$\overline{\Lambda}$  &309.68     &\bf 359.71    &85.32        &346.25        &418.10          &315.13      \\
\hline
$\Xi^0$               &39.12      &\bf 19.04     &29.25        &13.26         &21.84           &16.90       \\
\hline
$\overline{\Xi^0}$    &32.00      &\bf 15.49     &16.07        &11.16         &17.97           &13.60       \\
\hline
$\Xi^-$               &39.97      &\bf 19.45     &30.40        &13.54         &22.26           &17.30       \\
\hline
$\overline{\Xi^-}$    &32.00      &\bf 15.49     &16.07        &11.16         &17.97           &13.60       \\
\hline
$\Omega^-$            &19.79      &\bf 1.01      &8.03         &0.35          &1.17            &0.89        \\
\hline
$\overline{\Omega^-}$ &19.79      &\bf 1.01      &5.53         &0.35          &1.17            &0.89        \\
\hline
$\pi^+$               &928.62     &\bf 911.72    &435.26       &1015.69       &1043.13         &811.42      \\
\hline
$\pi^0$               &1877.62    &\bf 1843.35   &887.53       &2052.50       &2106.45         &1642.49     \\
\hline
$\pi^-$               &948.99     &\bf 931.63    &452.26       &1036.51       &1063.32         &831.07      \\
\hline
$K^+$                 &635.72     &\bf 686.80    &473.57       &630.45        &782.21          &613.97      \\
\hline
$K^-$                 &528.75     &\bf 570.90    &378.65       &534.03        &659.77          &503.06      \\
\hline
$K^0_L$               &589.21     &\bf 636.35    &435.36       &588.70        &728.55          &565.94      \\
\hline
$K^0_S$               &589.21     &\bf 636.35    &435.36       &588.70        &728.55          &565.94      \\
\hline
$\eta$                &162.89     &\bf 125.6     &358.21       &86.23         &144.64          &111.06      \\
\hline
\end{tabular}
\end{center}
\caption
{
The predicted hadronic observables in a central $Pb+Pb$ collision at 5.52 TeV/nucleon-pair CM energy.
}
\label{table_observ}
\end{table*}

In Table~\ref{table_ratios} we give all the abundance ratios which we used in Section~\ref{chap:app};
the data will show the qualities of the models in the last column.
The experiments and the references are identified, too. We have ignored individual data that seem to
be direct contradictions to data referred here. In case of 158 AGeV SPS we have not used
$K^+/\pi^+$ and in case of 130 AGeV RHIC we have not used $\Omega/\pi^-$, $K^-/\pi^-$,
$\overline{p} /\pi^-$ because we were not convinced that the pion background does not adulterate
the ratios. (Again, hadronisation into resonances being neglected)
In contrast, such ratios were kept for 200 AGeV RHIC because of the shortage
of ratios at this energy. If one does not like such ratios, one can use a derived
experimental ratio $K^-/\overline{p}$ instead of the last 2 rows given in the brackets.
The predictions of the models with least average errors are given, too: Model 9 for the
first 2 energies and Model 7 for RHIC energies.

\begin{table*}
\begin{center}
\begin{tabular}{|l|l|l|l|l|l|}
\hline
Energy/nucl. &Ratio               &Experiment    &Reference      &Measurement     &Model 9        \\
\hline
\hline
158 GeV SPS  &$\Xi^0/\Lambda$     &WA97          &\cite{let972}  &$0.14 \pm 0.02$ &$0.14 \pm 0.13$\\
\hline
             &$\overline{\Xi^0}/\overline{\Lambda}$
                                  &WA97          &\cite{let972}  &$0.26 \pm 0.05$ &$0.19 \pm 0.13$\\
\hline
             &$\Omega/\Xi^0$      &WA97          &\cite{let972}  &$0.19 \pm 0.04$ &$0.27 \pm 0.13$\\
\hline
             &$\overline{\Omega}/\overline{\Xi^0}$
                                  &WA97          &\cite{let972}  &$0.30 \pm 0.09$ &$0.35 \pm 0.13$\\
\hline
             &$\overline{\Lambda}/\Lambda$
                                  &WA97, NA49    &\cite{let972}  &$0.145 \pm 0.024$&$0.38 \pm 0.13$\\
\hline
             &$\overline{\Xi^-}/\Xi^-$
                                  &WA97          &\cite{let972}  &$0.27 \pm 0.05$ &$0.49 \pm 0.13$\\
\hline
             &$\overline{\Omega}/\Omega$
                                  &WA97          &\cite{let972}  &$0.42 \pm 0.12$ &$0.66 \pm 0.13$\\
\hline
             &$(\Xi^-+\overline{\Xi^-}) / (\Lambda + \overline{\Lambda})$
                                  &NA49          &\cite{let972}  &$0.13 \pm 0.03$ &$0.16 \pm 0.13$\\
\hline
             &$\overline{\Lambda} /  \overline{p}$
                                  &NA49          &\cite{let06}   &$1.05 \pm 0.16$ &$1.12 \pm 0.13$\\
\hline
             &$K^- /  K^+$        &NA49          &\cite{red02}   &$0.59 \pm 0.05$ &$0.77 \pm 0.13$\\
\hline
\hline
200 GeV SPS  &$\overline{\Lambda}/\overline{p}$
                                  &NA35          &\cite{let97}   &$0.80 \pm 0.25$ &$1.12 \pm 0.14$\\
\hline
             &$\Xi^-/\Lambda$     &WA85          &\cite{let96}   &$0.19 \pm 0.01$ &$0.15 \pm 0.14$\\
\hline
             &$\overline{\Xi^-}/\overline{\Lambda}$
                                  &WA85           &\cite{let96},\cite{let97}
                                                                 &$0.21 \pm 0.02$ &$0.19 \pm 0.14$\\
\hline
             &$(\Omega+\overline{\Omega}) / (\Xi^- + \overline{\Xi^-})$
                                  &NA35          &\cite{let97}   &$0.80 \pm 0.4 $ &$0.28 \pm 0.14$\\
\hline
             &$\overline{\Lambda}/\Lambda$
                                  &NA35          &\cite{bar90}   &$0.18 \pm 0.06$ &$0.38 \pm 0.14$\\
\hline
\hline
             &                    &              &               &                &Model 7        \\
\hline
\hline
130 GeV RHIC &$\overline{p}/p$
                                  &STAR          &\cite{red02}, \cite{red022}
                                                                 &$0.64 \pm 0.05$ &$0.59 \pm 0.06$\\
\hline
             &$\overline{\Omega}/\Omega$
                                  &STAR          &\cite{tom03}   &$1.00 \pm 0.2 $ &$1.00 \pm 0.06$\\
\hline
             &$\overline{\Lambda}/\Lambda$
                                  &STAR          &\cite{red02}, \cite{red022}
                                                                 &$0.77 \pm 0.05$ &$0.65 \pm 0.06$\\
\hline
             &$\overline{\Xi^-}/\Xi^-$
                                  &STAR          &\cite{red02},  \cite{red022}
                                                                &$0.81 \pm 0.05$  &$0.78 \pm 0.06$\\
\hline
             &$\pi^-/\pi^+$
                                  &PHOBOS        &\cite{red022}  &$0.95 \pm 0.06$ &$1.02 \pm 0.06$\\
\hline
             &$K^-/K^+$
                                  &STAR          &\cite{red022}  &$0.90 \pm 0.05$ &$0.82 \pm 0.06$\\
\hline
\hline
200 GeV RHIC &$\overline{p}/p$
                                  &PHOBOS, PHENIX&\cite{tom03}, \cite{bic04}
                                                                 &$0.84 \pm 0.04$ &$0.60 \pm 0.16$\\
             &                    &STAR, BRAHMS  &               &                & \\
\hline
             &$K^-/K^+$           &PHOBOS, PHENIX&\cite{tom03}, \cite{bic04}
                                                                 &$0.98 \pm 0.04$ &$0.82 \pm 0.16$\\
             &                    &BRAHMS        &               &                & \\
\hline
             &$\pi^-/\pi^+$
                                  &PHOBOS        &\cite{bic04}   &$1.02 \pm 0.02$ &$1.02 \pm 0.16$\\
\hline
             &$K^-/\pi^-$
                                  &STAR, BRAHMS  &\cite{tom03}   &$0.15 \pm 0.02$ &$0.61 \pm 0.16$\\
\hline
             &$\overline{p}/\pi^-$
                                  &BRAHMS        &\cite{tom03}   &$0.08 \pm 0.01$ &$0.16 \pm 0.16$\\
\hline
             &($K^-/\overline{p}$
                                  &STAR, BRAHMS  &\cite{tom03}   &$1.87 \pm 0.34$ &$3.81 \pm 0.16$)\\
\hline
\hline

\end{tabular}
\end{center}
\caption
{
Experimental and model particle ratios for different collision energies.
The errors of the penultime column are that of the ratio measurements.
The average errors of the models can be obtained from Table~\ref{table_errors} 
as: 0.13 for 158 GeV SPS, 0.14 for 200 GeV SPS, 0.06 for 130 GeV RHIC,
and 0.16 for 200 GeV RHIC.
}
\label{table_ratios}
\end{table*}

Now, let us compare methodically the measured and predicted ratios.
Generally, they must agree because the model errors were estimated in
Eq.~(\ref{eq:sigma2}) and Table~\ref{table_errors} from the differences. However, it is interesting
to see the possible individual big differences which would signal
shortcomings of even the best model of the original 1-10. Henceforth we
study the deviations $r_{i,obs} - r_{i,mod}$ whose errors are composed from the
individual measurement errors and the model error.
As for the error of the difference $\Delta$, its distribution is the convolution of those
of the r's, so the second central momentum is the sum of the constituents~\cite{janos}, i.e.
\begin{eqnarray}
&&      \sigma^2 = \sigma_1^2 + \sigma_2^2
\end{eqnarray}
but of course we do not know the distribution, only that it is nearer to
Gaussian than those of the r's.

Using the good rehadronisation model, the observed r's should fluctuate around
the model predictions, so the set ${\Delta_i}$ denoting individual ratios should
fluctuate around 0. Table~\ref{table_observ} contains 26 independent ratios.
Which of them 16 is not farther from 0 than 1$\sigma$. For the remaining 7 we get Table~\ref{table_sigmas}.
\begin{table*}
\begin{center}
\begin{tabular}{|l|l|l|l|l|l|}
\hline
Experiment  &Ratio     &$\Delta$\hspace{1.5cm} &$\sigma$\hspace{1.5cm} &Deviation/$\sigma$&Error integral  \\
\hline
158 GeV SPS &$\overline{\Lambda}/\Lambda$     &-0.235     &0.13          &1.85               &0.06            \\
\hline
158 GeV SPS &$\overline{\Xi^-}/\Xi^-$         &-0.22      &0.14          &1.57               &0.12            \\
\hline
158 GeV SPS &$\overline{\Omega}/\Omega$       &-0.24      &0.18          &1.33               &0.18            \\
\hline
158 GeV SPS &$K^- /  K^+$                     &-0.18      &0.14          &1.28               &0.20            \\
\hline
200 GeV SPS &$\overline{\Lambda}/\overline{p}$& 0.32      &0.29          &1.10               &0.27            \\
\hline
200 GeV SPS &$(\Omega+\overline{\Omega}) / (\Xi^- + \overline{\Xi^-})$
                                              & 0.52      &0.42          &1.24               &0.22            \\
\hline
200 GeV SPS &$\overline{\Lambda}/\Lambda$     &-0.20      &0.15          &1.33               &0.18            \\
\hline
130 GeV RHIC&$\overline{\Lambda}/\Lambda$     & 0.12      &0.08          &1.50               &0.13            \\
\hline
200 GeV RHIC&$\overline{p}/p$                 & 0.24      &0.16          &1.50               &0.13            \\
\hline
200 GeV RHIC&$K^-/\pi^-$                      &-0.46      &0.16          &2.88               &0.004           \\
\hline
\end{tabular}
\end{center}
\caption
{
10 ratios in Table~\ref{table_ratios} whose $\sigma$'s are above 1 
with one ratio whose $\sigma$'s is  above 2.
The difference $\Delta$ of columns 5 \& 6 is within 1$\sigma$ in 16 cases.
}
\label{table_sigmas}
\end{table*}

There was really no measurement for the synthetic last bracketed ratio at 200 GeV RHIC,
so that value is not analysed here. The last column is the error integral
\begin{eqnarray}
        1-erf(n) = \sqrt{2/\pi}   \int_n^{\infty} \! e^{-u^2/2} \mathrm{d} u
\end{eqnarray}
Were the distribution of $\Delta$ Gaussian, the last column would show the probability of such
a random fluctuation; it is at least an estimation for it. The position is the position
within the actual group in Table~\ref{table_ratios}.

Individual measurements up to 2$\sigma$ are generally not regarded as too problematic.
So 9 out of the 10 ratios outside 1$\sigma$ are not necessarily signals of problems. $\chi^2$
tests might be advisable, but will not be done here. However, the $K^-/\pi^-$  ratio
at 200 GeV RHIC is surely problematic.

Our guess is that we see consequences of enhanced pion yields, and it is probably
enhanced by rehadronisation into resonances. As discussed already in Section~\ref{chap:numr}, Point 2,
the encounter of quarks may result in resonances instead of stables. Were the timescale
of phase transition significantly longer than the resonance lifetimes, the quarks could
try again. However, both timescales are $10^{-23}$ s, so the two processes do not clearly separate;
some resonances may decay in the hadronic phase, giving extra mesons, mainly $\pi$'s.

We are not really manufacturing a full Model 12 including resonances; a lot of presently badly
known new parameters should be used for that. But just for demonstration consider
the case when the final baryon yields are unchanged, but half of them come through
the corresponding member of the decuplet. (Of course, $\Omega$ is an exception, being a stable
and, in the same time, a member of the decuplet.) Then 3 model predictions of the 5
200 GeV RHIC yields in Table~\ref{table_observ} are unchanged, but 2 gives new results.
For estimation take the model errors equal to that of Model 7; then we get Table~\ref{table_sigmas2}.
That is our "Model 12"; The quotation marks are used because it not a fully elaborated model but
rather an estimation.
\begin{table*}
\begin{center}
\begin{tabular}{|l|l|l|l|l|}
\hline
Experiment  &Ratio                &$\Delta$   &$\sigma$      &Deviation/$\sigma$ \\
\hline
200 GeV RHIC&$K^-/\pi^-$          &-0.11      &0.16          &0.69               \\
\hline
200 GeV RHIC&$\overline{p}/\pi^-$ & 0.09      &0.16          &0.57               \\
\hline
\end{tabular}
\end{center}
\caption
{
The so far problematic ratios at 200 GeV RHIC in "Model 12".
}
\label{table_sigmas2}
\end{table*}
Both problematical ratios have got within 1$\sigma$ to 0.

So surely there are stables which resulted through resonances. Surely pion numbers
will be then higher than in the present models. However, kaon numbers would be unchanged
in first approximation since the decays mainly go as $X^* \rightarrow X\pi$.
So either we should handle the resonances properly before predictions are made;
or as a first step we should ignore any ratio involving pions. As a second step,
the $\pi$ surplus will be between 0 and the sum of the numbers of 1/2-spin baryons + antibaryons.

The synthetic ratio $K^-/\overline{p}$ is still rather high from Model 7 compared to the ratios
of 2 measured ratios. However, the statistics of such derived quantities need serious cautions.

We emphasize that an almost infinite number of different models can reproduce more or less
the yields published so far. Our models are focused on rehadronisation, but they were not too
interested about, say, collective phenomena in the QG or hadron phase. Such collective phenomena
may cause quite remarkable effects, as e.g.~\cite{verte} can predict very enhanced $\eta$'
yields, and then $\pi$ production, via the mass of $\eta$' meson much lower than the free mass.
Such effects are actually quite importans.
However, note that $\eta$', while it is technically a stable, is rather a short lived particle
with $10^{-21}$s lifetime and the situation is unfamiliar as in~\cite{luk86} shown many years
ago that even existing particleis may or may not form an independent thermodynamic degree of
freedom. In addition, lots of formally different approches may result in
similar yields, as it was shown in~\cite{luk86}.
At the present status of knowledge we rather want to compare as many models as possible
and we are not looking for the best one.

At the end of this Section it is the proper place to note that there are very preliminary 
yields from the new range about $E_{CM}$ = 2.76 TeV. E.g.~\cite{mulle} gives 2*4 ratios for 
nearly central collisions, without error bars. For antiparticles the ratios are almost 
exactly the same as for particles, and, according to what we mentioned above, it is better 
to leave out the pions. Then we remain with three ratios, $K^+/p$, $\Xi^-/p$ and $\Omega/p$. 
Comparing these very approximate ratios with predictions of Model 7 for 0.2 \& 5.52 TeVs 
we get Table~\ref{table_measure}.
\begin{table*}
\begin{center}
\begin{tabular}{|l|l|l|l|}
\hline
Ratio       &0.2 TeV, Model 7     &2.76 TeV, prelimininary &5.52 TeV, Model 7  \\
            &                     &measurement             &                   \\
\hline
$K^+/p$     &2.8   $\pm$ 0.34     &3.2                     &2.9 $\pm$ 0.49     \\
\hline
$\Xi^-/p$   &0.078 $\pm$ 0.013    &0.12                    &0.083 $\pm$ 0.014  \\
\hline
$\Omega/p$ &0.004  $\pm$ 0.0007   &0.02                    &0.004 $\pm$ 0.0007 \\
\hline
\end{tabular}
\end{center}
\caption
{
}
Comparison to non-pionic preliminary ratios for 2.76 TeV in~\cite{mulle}. The errors
of the model calculatios are based on Table~\ref{table_observ}. 
\label{table_measure}
\end{table*}

While the first two ratios would be within error after the manner of Table 7, 
$\Omega/p$ of Model 7 is too big, by cca. a factor of 5. However, Model 5 would 
give a ratio even bigger by almost 8. This again would suggest something between
 Models 5 \& 7 as it was seen in Section~\ref{chap:app} at RHIC energies.

\section{Conclusion}
\label{chap:con}

At the end we should answer the question of QG phase. In some sense we have done that:
we predict the yields of our Model 7, with S/N=65.7, Column 3 of Table~\ref{table_observ}.
Now, Model 7 contains a phase transformation into QGP, then a definite amount of gluon
fragmentation, and then a rehadronisation according to some rules. So if the yield ratios will
be near to those calculable from our preferred Column 3, then indeed the detected ratios confirm
the transient existence of the QGP. However, the old story of too much $K^+$ in
1987~\cite{plos,tann,luk900} is a warning that even a good agreement with
the predicted ratios would not be a proof. A proof would be almost impossible.
Observe that, e.g., a completely hadronic explanation of "too much" $K^+$ at 14.5 GeV fixed
target could be "simply" an enhanced rate of the reaction
\begin{eqnarray}
&&      p + p \rightarrow p + K^+ + \Lambda
\label{eq:pp}
\end{eqnarray}
above T=150 MeV.

Namely, the existence of a QG phase is something with definite meaning only in thermodynamic
context; and~\cite{lan61} gives the rigorous prescription to decide if a definite
thermodynamic phase is present. We cannot perform the needed operations in the later
hadronic phase. In principle leptons may originate in QGP and so carry some direct
information; but that is beyond the scope of the present paper.

As told earlier, our preferred prediction is Column 3 of Table~\ref{table_observ}. You can form
then the actual ratios measured in the future. (Corrections for the slight $d$ to $u$ surplus
at initial condition have been done in Table~\ref{table_observ}.)
The other columns are partly for comparison, partly for cases when "something unexpected happens"
between TEVATRON and LHC energies.

We have seen that at RHIC energies the performance of our Model 5 was only slightly worse that
of Model 7; but combination of the two models did not help, so Model 7 was chosen as the absolute best.
Now, Model 5, without the selecting mechanism of hadronic final state data, results in more
heavy hyperons than Model 7 (in $\Omega$ the increase is almost 20-fold), and therefore, slightly less
lighter ones. It is not quite impossible that with increasing energy the combinations feel less
the final state masses; not because of higher energy (entropy would not increase much) but
because less time for new trials in hadron formation; but still we believe rather in Model 7.
But data of \cite{mulle}  suggest that the model might suppress $\Omega$ too strongly, while
Model 5 does not do it sufficiently.

Model 11, Model 7 with enhanced gluon fragmentation, would be observed, according to
Column 5, as slight increases in lighter antihadrons and slight decreases in heavier ones.
The differences are moderate, and it is questionable if the observations will be able to
distinguish Models 7 \& 11. Yield ratios nearer to Model 11 than to Model 7 would not automatically
suggest enhanced gluon fragmentation; rehadronisation into resonances and then decay in the
hadronic matter leads to a $\pi$ surplus which is also true at enhanced gluon fragmentation.
On the other hand, it seems that higher S/N (last column in Table~\ref{table_observ})
would be distinguishable from resonance decay. Anyway, $K^-/\pi^-$ and $\overline{p}/\pi^-$,
we think, again could signal resonance decays in hadronic phase if the ratios were much smaller
than any columns of Table~\ref{table_observ}. "Model 12", as told, is not a full model but an
estimation for the pion surplus from resonances. We simply assumed that during the hadronisation
half of the emerging baryons were stables and the other halp resonances. The resonances decay
in the hadronic phase. This "Model" has been explained at the end of Section~\ref{chap:dis}.

Column 6 shows the outputs for Model 7 at a higher S/N=75. The entropy would increase if
something unexpected happens with the cross sections between TEVATRON and LHC energies.
For ratios the differences between Columns 3 \& 6 are slight, because all particle yields
increase in Column 6. Maybe the difference would be observable in the antiparticle/particle ratios;
they are higher in Column 6, but the difference is not greater than 3 \%.

And finally, let us see the famous $K^+/\pi^+$ ratio which provoked discussions back to 1987
about the formation of QGP~\cite{plos,tann,luk900}; which lead to the ALCOR model
in 1999~\cite{zim00}, and which was in early times in general expected a good QGP signal.
For comparison we give $K^-/\pi^-$ in Table~\ref{table_K_pi}, as well.
\begin{table*}
\begin{center}
\begin{tabular}{|l|l|l|l|l|l|l|}
\hline
            & Year-1987     &Model 5      &Model 7       &Model 9     &Model 11         &Model 7    \\
\hline
Note        & 14.5 GeV fixed t.&fragmentation&favourite     &hadronic    &enh.fragmentation&S/N=75 \\
\hline
$K^+/\pi^+$ &  0.24         &0.685        &0.753         &1.088       &0.621            &0.750      \\
\hline
$K^-/\pi^-$ &$\approx$0     &0.557        &0.613         &0.837       &0.515            &0.620      \\
\hline
\end{tabular}
\end{center}
\caption
{
 $K/\pi$ ratios in Pb+Pb reactions at 5.52 TeV.
}
\label{table_K_pi}
\end{table*}

Now, the morale is clear but somewhat disturbing. At "low energies" a $K/\pi$ ratio much higher
for kaons than for antikaons had been believed to be a quark signal.
But for 5.52 TeV all models give higher $K^+/\pi^+$ than $K^-/\pi^-$ ratios;
and the highest ratios are given by Model 9 completely without QGP!
(See also the $K/\pi$ discussion in~\cite{holba})

One possible conclusion is that great care is needed when looking for signals. Another is that
maybe at high energies the situation differs from those at lower ones. Both conclusions are indeed
probable, even if not quite quantitative. However, there is something more, to which
Table~\ref{table_K_pi} is a warning.

Models 1-8 \& 10-11 contain QGP, with quarks in equilibria within (probably a good
assumption, reaction times being guessed short in QGP); the rehadronisation assumptions
differ (and they may be not equilibrium ones). The completely hadronic Model 9 contains
hadrons in equilibria.

Now, for the QGP quark to antiquark surplus is necessary because of initial condition,
except for s, where $\overline{s}/s$=1 because all strange quarks were produced in QGP
and strangeness is conserved.

Being $K^+ = (u\overline{s})$ and $K^- = (\overline{u}s)$, the ratio $K^+/K^->$ 1 is
nontrivial from quark composition. However, note that all models with QGP feel the initial conditions, so produ
ce
baryon to antibaryon surplus. Then $K^+/K^->$ 1 is a necessary consequence of conservations.

As for Model 9 without QGP it is an equilibrium model. If time were long enough,
high $K/\pi$ ratios would be obtained. Now, in 1987 the time estimated seemed not enough
for this, and we still believe this, because $K^-/\pi^-$ was cca. 0~\cite{tann}.
Namely, consider initial N+N collisions. The simplest result is 3 particles, say N, a hyperon
and a kaon. Antikaon instead of kaon is impossible for 3 particles because of
strangeness conservation. Therefore, reactions resulting in $K^-$ would have low rates anyways;
although the energy barrier is much lower in QGP for $qqq \rightarrow qqqs\overline{s}$
then in HM for $NN \rightarrow NK\Lambda$, cca. 300 MeV in the first case and almost
700 MeV in the second.
So $K^-/K^+ <<$ 1 in 1987 in itself was not a signal either for or against liberated quarks,
only against quarks in complete equilibrium; a "too high" $K^+$ number suggested liberated
quarks, but because of $K^-/K^+ <<$ 1 is not a "quite proper" QGP.

It seems that at RHIC energy, S/N=58 a proper QGP phase was formed. We cannot prove this
being the QGP unreachable for us; but the hadronic yields prefer QGP.
Then we have no reason to doubt in QGP at 5.52 TeV.

\paragraph*{Acknowledgements}
We think formal acknowledgements are unnecessary to the colleagues
referred below. J. Zim\'anyi triggered the studies and gave lots of
impetus for us.
H-W. Barz, T. S. Bir\'o, L. P. Csernai, T. Cs\"org\H{o}, B. Jakobsson,
B. K\"ampfer, Gy. Kuti, P. L\'evai, K. Martin\'as, L. Polonyi, K. Szlach\'anyi, Gy. Wolf
were not only coauthors of one of the present author in works cited here,
but also they worked in producing/improving the models used here.
Thanks go to G\"osta Gustafson of University of Lund for valuable discussions.
The authors gratefully acknowledge the support of the Hungarian OTKA grant NK 101438.

\begin{appendix}

\paragraph*{Appendix}
\def\theequation{\thesection.\arabic{equation}}
\let\oldsection\section
\renewcommand{\section}[1]{\setcounter{equation}{0} \oldsection{#1}}

\section{List of the used rehadronisation models}
\label{app:mod}

Here we use 12 models, that, however, are practically 8 at LHC energies. They are as follows.
Models from 1 to 8 are rehadronisation model in $2^3$ arrangement, as a binary number + 1. The
models were discussed in Ref.~\cite{luk91} \& \cite{luk96}. The conditions for the states of the 3 
digits are as follows:

\begin{enumerate}
\renewcommand{\labelenumi}{\arabic{enumi})}
\item Final state compressibilty is involved (1) or not (0). On this digit, however,
      dependence is negligible at 2.76+2.76 TeV.
\item Final state hadronic masses are taken into account (1) or not (0). This influences 
      the yields and the best model cannot be selected a priori because the rehadronisation
      is not an equilibrium process.
\item Gluon fragmentation takes place in QGP (1) or it does not (0). If it does, we take its
      parameters from Ref.~\cite{koc86}.
\end{enumerate}
So Models 1-8 refer to the 3 digit states "000"-"111" according to Table~\ref{table_mod8}, in summary.
\begin{table}[h]
\begin{tabular}{|c|c|c|c|}
\hline
Model        & Gluon               & Hadronic           & Final state          \\
\#           & fragmentation       & masses             & compressibility      \\
\hline
    1        & 0                   & 0                  & 0                   \\
    2        & 0                   & 0                  & 1                   \\
    3        & 0                   & 1                  & 0                   \\
    4        & 0                   & 1                  & 1                   \\
    5        & 1                   & 0                  & 0                   \\
    6        & 1                   & 0                  & 1                   \\
    7        & 1                   & 1                  & 0                   \\
    8        & 1                   & 1                  & 1                   \\
\hline
\end{tabular}
\caption
{ A group of rehadronisation models including the 3 options in columns 2-4. }
\label{table_mod8}
\end{table}

Model 9 is a purely hadronic phase one, and Model 10 is a sequential fission one, see 
Refs.~\cite{luk91} \& \cite{luk96}.

Model 11 is a version of Model 7, but with gluon fragmentation enhanced to the energetically
possible maximum, see Ref.~\cite{zim93} \& \cite{kfki93}.

Model 12 is not an entirely elaborated model, but an educated guess, when everything starts 
as in Model 7, but half of the resulting baryons enter the hadronic phase as resosances and
the decay there, giving surplus mesons, mainly $\pi$'s.

\section{Transformation to partially specific variables}
\label{app:tra}

Eq.~(\ref{eq:extr}-\ref{eq:hats}) guarantee that we can calculate functions from $\hat{s}$.
Now, let us use $\hat{s}$.

As told Uneq.~(\ref{eq:s0}) must hold, for any timelike $u^i$ field. In $\hat{s}_{;r}$ the derivatives
$n_{,r} u^r$ and $e_{,r} u^r$ can be substituted from Eqs.~(\ref{eq:T0},\ref{eq:n0})
and then we get
\begin{eqnarray}
&&     \hat{s},_t Dt + (\hat{s} - n \hat{s},_n - e \hat{s},_e - k \hat{s},_e) u^r;_r 
       + t^r(z,_r - \hat{s},_e \beta,_r) \nonumber \\
&&     + t^r;_r (z - \hat{s},_e \beta)
       - \hat{s},_e \beta t^r u^s u^{r;s} - \hat{s},_e (\gamma - k / t^2) t^r t^s u_{r;s} \nonumber \\
&&     \geq  0 \nonumber \\
&&     D \equiv u^s \partial_s
\label{eq:appa}
\end{eqnarray}
identically i.e. for any u; and several terms apppear with quite various u-dependences.
Collecting the terms with the same u-dependences, Uneq.~(\ref{eq:s0}) is guaranteed only if the terms
of different u-dependences satisfy equations independently. That is set in Eq.~\ref{eq:uneqs}).

\section{Evaluation of $\lambda$ of ultrarelativistic collisions}
\label{app:eva}

This Appendix is necessary to avoid inconsistencies in the main text. Namely,
in this paper t stands for 3 quite different quantities: time, the specific extra
extensive and one invariant familiar in 2-particle collisions. Similarly, s
means both entropy density and another invariant, while $u$ is both the velocity
vector and the third invariant. So we have separated the part most confusing
and here we try to be careful and explicit in the formulation.

Consider an elastic two-body collision where the incoming two particles have
the original momenta of the beams. By appropriate choice of the coordinate
system and by momentum conservation we always can write:
\begin{eqnarray}
&& u_{1/2}^i  = \{(1+v^2)^{1/2}, \pm v, 0, 0\}  \nonumber \\
&& u_{3/4}^i  = \{(1+v^2)^{1/2}, \pm (v^2-w^2)^{1/2}, \pm w, 0 \}
\label{eq:appb1}
\end{eqnarray}
Now, the collision is characterised by a triad (s,t,u) which here for clarity
will be written in {\bf bold}. They are defined as
\begin{eqnarray}
&& {\bf s}  = (p_1+p_2)^2 \nonumber \\
&& {\bf t}  = (p_1-p_3)^2 \nonumber \\
&& {\bf u}  = (p_1-p_4)^2
\label{eq:appb2}
\end{eqnarray}
but here we will not use $\bf u$. Using Eqs.~(\ref{eq:appb1},\ref{eq:appb2}) we get
\begin{eqnarray}
&& {\bf s}  = 4m^2(1+v^2) \\
&& {\bf t}  = -2m^2v(v-(v^2-w^2)^{1/2})
\label{eq:appb3}
\end{eqnarray}
Expressing it with the decrease of $v, (v^2-w^2) \equiv v-\Delta$, it is simply
\begin{eqnarray}
&& \Delta  = {\bf t}/2m^2c^2 v
\label{eq:appb4}
\end{eqnarray}
The average loss of the longitudinal momentum per collision is then
\begin{eqnarray}
&& <\Delta>  = (1/2m^2c^2 v)(\int{{\bf t} \sigma({\bf t}) d{\bf t}} / \int{\sigma({\bf t}) d{\bf t}} )
\label{eq:appb5}
\end{eqnarray}
where v can be substituted by s via Eq.~(\ref{eq:appb5}).

Since $\lambda$ is the source term of q, the density of the momentum-like extra extensive,
$\lambda$ must be the product of 3 terms: the collisions/time calculated from the total
cross section, the average momentum loss in one collision (this is $<\Delta>$),
and the actual density. The last is cca. $2n_0$. As for the $s$ dependence we used the
almost $ln^2 s$ curve of~\cite{islam} and~\cite{ferro}.

\section{The $\Lambda$-$\Sigma$ sector}
\label{app:the}

Among the stables we find 4 hyperons with ($qqs$) quark composition:
($uus$) is $\Sigma^+$, ($dds$) is $\Sigma^-$, but ($uds$) may both be $\Sigma^0$ and $\Lambda$.
The difference is in the relative spin positions. Measurements indicate that ($qq$) is parallel
in $\Sigma$ and antiparallel in $\Lambda$; and the antiparallel configuration is absent for
($uu$) and ($dd$) (For $\Sigma$ $J^P$ is $1(\frac{1}{2}^+)$, for $\Lambda$ $J^P$ is $0(\frac{1}{2}^+)$ ).
No doubt, we see a consequence of the Pauli principle.

Of course, in the spin-3/2 decuplet antiparallel spin position is per definitionem
impossible, so there only a triplet $\Sigma^*$ appears with $J^P = 1(\frac{3}{2}^+)$.
The interesting fact is the 77 MeV
mass difference between $\Lambda$ and $\Sigma^0$; $\Lambda$ is preferred. Obviously this comes from the spin
positions, but our rehadronisation models neglected the energy difference within
the sectors. But then the yields amongst the Y's are debatable from purely the quark
numbers, while the total Y yield is unique, and the differences do not influence the
other baryons. We think that the final answer would need the handling of resonances in
rehadronisation, and in this moment that would delay the predictions. So instead we
elaborated 2 simple models and chose that which kept charge conservation better.
Here we discuss only the charge-symmetric distinction of $\Lambda$ and $\Sigma$; afterwards the
inclusion of u/d $\neq$ 1 is as for the other particles.

Proposal 1 was $\Sigma$=2*Y/3,  $\Lambda$=Y/3. This assumes that every initial spin position remains
if a stable is possible, and fades away if not. So $\Sigma^{\pm}$ is Y/6 and $\Sigma^0$ = $\Lambda$ =Y/3.

Proposal 2 is $\Sigma$ = $\Lambda$ =Y/2, the parallel and antiparallel positions of the non-strange
quarks are equally probable. Then the charged pre-$\Lambda$'s first form $\Sigma^*$'s,
which then decay into $\Lambda$ and $\pi$'s.

While Proposal 1 seemed more logical, it generally resulted in Z $\approx$ 174, while
Proposal 2 in  Z $\approx$ 172. Exact charge conservation would be Z=164.
Probably,  $\Lambda$ $>$ $\Sigma^0$ because the $\Lambda$ mass is less;
but this would not help the Z conservation,  and here for now we stop.

\end{appendix}

\end{document}